\documentclass[twocolumn,twocolappendix]{aastex63}
\usepackage[version=4]{mhchem}
\usepackage{physics}
\usepackage{comment}
\usepackage{amsmath}
\usepackage{CJK}
\usepackage{mathtools}
\usepackage[version=4]{mhchem}

\newcommand{\E}{E_{r}}
\newcommand{\F}{\vec{F}_{r}}
\newcommand{\eg}{e_{\text{g}}}
\newcommand{\kp}{\kappa_{\rm P}}
\newcommand{\ke}{\kappa_{\rm e}}
\newcommand{\kr}{\kappa_{\rm R}}
\newcommand{\cmsg}{cm$^{2}\cdot$ g$^{-1}$}
\newcommand{\gcmc}{g$\cdot$cm$^{-3}$}
\newcommand{\ergcmc}{erg$\cdot$cm$^{-3}$}
\newcommand{\cms}{cm$\cdot$s$^{-1}$}
\newcommand{\kms}{km$\cdot$s$^{-1}$}

\newcommand{\half}{\frac{1}{2}}
\newcommand{\tg}{T_{\text{g}}}
\newcommand{\trad}{T_{\text{rad}}}
\newcommand{\er}{\epsilon_{r}}
\newcommand{\sig}{\text{sig}}

\submitjournal{ApJS}

\shorttitle{Guangqi: A 2D radiation hydrodynamic code}
\shortauthors{Chen \& Bai}
\graphicspath{{./}{figures/}}
\begin{document}
\begin{CJK*}{UTF8}{gbsn}
\title{Guangqi: A two-dimensional radiation hydrodynamic code with realistic equation of states}

\correspondingauthor{Zhuo Chen (陈卓)}
\email{chenzhuo\_astro@mail.tsinghua.edu.cn}\\
\correspondingauthor{Xue-Ning Bai (白雪宁)}
\email{xbai@mail.tsinghua.edu.cn}

\author[0000-0002-0786-7307]{Zhuo Chen(陈卓)}
\affiliation{Institute for Advanced Study, Tsinghua University \\
Beijing 100084, China}

\author[0000-0001-6906-9549]{Xue-Ning Bai (白雪宁)}
\affiliation{Institute for Advanced Study, Tsinghua University \\
Beijing 100084, China}
\affiliation{Department of Astronomy, Tsinghua University \\
Beijing 100084, China}

\begin{abstract}
We present {\tt Guangqi}, a new two-dimensional, finite-volume radiation hydrodynamics code designed for high-performance astrophysical simulations. The code simultaneously resolves the hydrodynamic equations for complex equations of state (EoS) and implicit radiation transport under the flux-limited diffusion approximation. Written in Fortran and parallelized via the Message Passing Interface. {\tt Guangqi} supports analytic hydrogen and helium EoS under the assumption of local thermal and chemical equilibrium. The framework is compatible with both Cartesian and spherical-polar geometries---utilizing non-uniform grid spacing---and incorporates static (SMR) and adaptive mesh refinement to optimize computational efficiency. To address the inherent challenges of angular momentum conservation in spherical-polar coordinates, we implement a robust and consistent ``passive scalar angular momentum algorithm" (PSAMA). Domain decomposition is managed through both Z-order and Hilbert space-filling curves to ensure scalability. The code has been rigorously verified against a suite of standard benchmarks and newly designed test cases specifically intended to diagnose the non-linear coupling between gas dynamics, intricate EoS, radiation transport, and angular momentum conservation.
\end{abstract}



\section{Introduction} \label{sec:intro}

Astrophysical fluids are inherently multiphysics, multiscale, and highly nonlinear. They often span vast dynamical ranges in density and temperature, involving non-ideal processes such as phase transitions (e.g., ionization and dissociation). This complexity presents a significant modeling challenge, as the resulting equations of state (EoS) become highly intricate and non-linear. Furthermore, treating phase transitions across diverse physical conditions requires a rigorous thermodynamic framework, which is often coupled with radiation transport---introducing further computational complications. Examples of such phenomena are ubiquitous, including type II supernovae (SNe) \citep{popov1993,smith2017,fang2025}, dwarf novae \citep{lasota2001,hameury2020}, luminous red novae (LRNe) \citep{tylenda2011,chen2024,chen2025}, common envelope evolution (CEE) \citep{ivanova2013,moreno2022,vetter2024}, and star/planetary formation \citep{larson1969,bhandare2018,chen2022}. Despite their importance, there remains a lack of multi-dimensional computational tools capable of simultaneously handling complex EoS and radiation transport. This deficiency motivates the development of a new astrophysical fluid dynamics code to bridge this gap.

We begin with the equation of radiative transfer---a 6D integro-differential equation dependent on spatial coordinates, angles, and frequency \citep{mihalas1984,castor2004}. It is customary to take angular moments of this equation to reduce its dimensionality. The simplest model for radiation transport involves taking the zeroth moment under the flux-limited-diffusion (FLD) approximation \citep{levermore1981}. FLD assumes a closure relation between radiation flux and radiation energy that transitions from diffusion to free-streaming between the optically thick and thin regimes. While the FLD equation generally requires implicit solvers (see Section \ref{sec:rad} for details), it has been widely adopted in various radiation hydrodynamic (RHD) codes \cite[e.g.][]{kolb2013,ramsey2015,colombo2019a}. However, a primary drawback of the FLD approximation is its inability to properly capture the anisotropy of radiation transport (e.g., shadows) and its potential for inaccuracy at intermediate optical depths \citep{davis2014}.

Improvements can be achieved by integrating the radiative transfer equation through both the zeroth and first moments, closing the system with the M1 approximation \citep{levermore1984,dubroca1999}. This two-moment formulation is hyperbolic, with characteristic wave speeds approaching the speed of light \citep{skinner2013}, which can impose stringent constraints on the timestep. Consequently, the M1 approximation is often paired with the reduced speed of light approximation (RSLA) \citep{gnedin2001,skinner2013} and an explicit solver. However, the RSLA can violate energy conservation \citep{fuksman2021,wibking2022}. While the M1 approximation effectively captures shadow effects \citep{gonzalez2007,skinner2013,kannan2019,bloch2021,fuksman2021,wibking2022}, it fails in ``cross-beam" scenarios \citep{jiang2012,jiang2021} because the closure relation cannot represent beams originating from multiple directions simultaneously.

The most accurate approach involves directly solving the radiative transfer equation across a discrete set of solid angles. This method provides superior local closure relations between the radiation pressure tensor and radiation energy density, leading to the variable Eddington tensor (VET) method \citep{davis2012,jiang2012}. Recently, it has become feasible to evolve the full radiative transfer equation coupled with hydrodynamics \citep{jiang2021,ma2025}. The primary disadvantage of these approaches is their high computational cost compared to FLD or M1 approximations; maintaining high precision of energy conservation often requires additional iterations, further reducing computational efficiency.

Regardless of the approximation used, proper treatment of the coupling between radiation and gas energy is essential for conservation. This coupling is typically so strong that the thermodynamic equations become stiff ordinary differential equations (ODEs) \citep{colombo2019a} requiring implicit solutions. Furthermore, a complex EoS can become highly nonlinear as gas temperatures fall in the range of phase transitions, complicating the implicit solvers. Because local thermodynamics are also coupled to non-local radiation transport, the system is typically solved via a fully implicit scheme \citep{kolb2013} or an implicit-explicit (IMEX) scheme \citep{wibking2022}, depending on the transport solver.

In this work, we introduce {\tt Guangqi} (光启)\footnote{We choose the name as a tribute to Xu Guangqi (徐光启), a mathematician and astronomer who collaborated with Matteo Ricci; the name also translates to ``enlightened by light" in Chinese.}, a new two-dimensional (2D) grid-based RHD code designed to handle general EoS. {\tt Guangqi} is a finite-volume Godunov code compatible with Cartesian and non-uniform spherical polar coordinates; in the latter, it is specifically designed to precisely conserve angular momentum. To maintain high accuracy in energy conservation while minimizing computational overhead, we adopt the FLD approximation. The equations are coupled with a general EoS, and the system is solved fully implicitly. To validate the code, we have devised a suite of numerical tests that solve RHD and EoS concurrently for diagnostic purposes. Currently, {\tt Guangqi} is configured to solve the RHD equations using a hydrogen (\ce{H2}, \ce{H}, \ce{H+}, and \ce{e-}) and helium (\ce{He}, \ce{He+}, \ce{He^{2+}}, and \ce{e-}) EoS \citep{chen2024} based on the Saha equations, assuming local thermodynamic equilibrium (LTE). We have also implemented adaptive mesh refinement (AMR) to facilitate simulations with large dynamical ranges. This new 2D code is ideally suited for long-term axisymmetric astrophysical problems involving complex thermodynamics and optically thick regions, such as CEE, LRNe, SNe, and star/planet formation. A new simple code can also offer an ideal testbed for novel algorithms, such as the radiation hydrodynamics with a complex EoS and the ``passive scalar angular momentum algorithm" (PSAMA) that will be presented in this paper.

The remainder of this paper is organized as follows. Section \ref{sec:physics} details the relevant physics and governing equations. Section \ref{sec:method} describes our numerical methodology. Section \ref{sec:framework} introduces the domain decomposition strategy for parallelization. We present numerical test results and comparisons against known solutions in Section \ref{sec:tests}. Finally, we provide a summary and concluding remarks in Section \ref{sec:disscusions}.

\section{The General Formalism} \label{sec:physics}

{\tt Guangqi} solves the 2D time-dependent radiation hydrodynamic equations with a general EoS.
The RHD governing equations in the conservative form are,
\begin{eqnarray}
	\pdv{\rho}{t}+\nabla\cdot(\rho \vec{v})&=&0,	\label{eqn:mass1} \\
    \pdv{(\rho\vec{v})}{t}+\nabla\cdot(\rho\vec{v}\vec{v}+p\mathbf{I})&=&\rho\vec{f},    \\
    \pdv{E}{t}+\nabla\cdot[\vec{v}(E+p)]&=&G^{0}+S_{\text{g}}+\rho \vec{f}\cdot\vec{v}, \label{eqn:gasenergy}        \\
    \pdv{\E}{t}+\nabla\cdot(\E\vec{v})+\nabla\cdot\F&=&-G^{0}, \label{eqn:radtrans}
\end{eqnarray}
where $\rho, \vec{v}, E, p\mathbf{I}$, and $\vec{f}$ are the density, velocity, gas total energy density, pressure tensor, and external force such as gravity), respectively. The total energy includes the kinetic energy and internal energy,
\begin{eqnarray}
	E&=&\rho(\frac{v^{2}}{2}+\epsilon),	\\
	\epsilon&=&\epsilon(\rho,\tg),	\label{eqn:eos}
\end{eqnarray}
where $v,\epsilon,\tg$ are the speed, specific internal energy, and gas temperature, $S_{\text{g}}$ represents external heating or cooling source terms (e.g., irradiation).
For the gas consisting of $k$ species, with the number density of each species $i$ being $n_i$, and the mass of each gas molecule/atom being $m_i$, we simply have,
\begin{eqnarray}
    \rho&=&\sum_{i=1}^{k}n_im_i, \\
    p&=&\sum_{i=1}^{k}n_{i}k_{b}\tg,
\end{eqnarray} 
where $k_b$ is the Boltzmann constant, and the pressure $p$ is given by the ideal gas law. Equation \ref{eqn:eos} is a general EoS and can be nonlinear for realistic gas, where we assume that all the species are in LTE and $n_{i}$ can be solved with Saha equations. \cite{chen2019} introduced a fast and accurate EoS solver to solve the Saha equations of hydrogen species, which we adopt in this work.

The radiation-related quantities are the radiation energy density $\E$, and radiation flux $\F$, measured in the lab frame, and the radiation-matter coupling strength $G^{0}$. They are defined as,
\begin{eqnarray}
	\E&=&\frac{1}{c}\int I_{\nu}d\omega d\nu,\label{eqn:eraddef}\\
	\F&=&\int I_{\nu}\mathbf{\hat{\Omega}}d\omega d\nu,\\
        G^{0}&=&\rho c\kp\E-\eta,	\label{eqn:radsource}
\end{eqnarray}
where $\kp$ and $\eta$ are the Planck mean opacity and emissivity measured in the co-moving frame, $\omega$, $\mathbf{\hat{\Omega}}$, and $\nu$ are the solid angle, unit vector in the direction of the solid angle, and frequency. The specific intensity $I_{\nu}$ is the fundamental quantity in the radiative transfer which depends on the coordinate, solid angle, and frequency. Its value can be obtained by solving the radiative transfer equation \citep{mihalas1984,castor2004}, which can be quite complicated. For convenience, we can equivalently express radiation energy density in terms of an effective radiation temperature (which is rigorous when radiation is isotropic and black body),
\begin{equation}
	\E=a_{R}\trad^{4},
\end{equation}
where $a_{R}$ is the radiation constant.

To calculate the radiation flux $\F$, we employ the FLD approximation for simplicity, given by \citep{levermore1981}
\begin{equation}
	\F=\frac{-c\lambda}{\kr\rho}\nabla\E\equiv-D\nabla\E	\\
\end{equation}
where $D$ is the diffusion coefficient, 
$\kappa_R$ is the Rossland mean opacity. 
In the FLD approximation, $\lambda$ is the flux limiter which should have the following asymptotic behaviors,
\begin{equation}\label{eqn:asymptotic}
\frac{c\lambda}{\kr\rho}\nabla\E\rightarrow\begin{cases}
	\frac{c}{3\kr\rho}\nabla\E&\quad\quad\rm{optically\ thick	}\\
	c\E &\quad\quad\rm{optically\ thin}\\
\end{cases}
\end{equation}
In this paper, the flux limiter we choose is based on \cite{minerbo1978},
\begin{equation}\label{eqn:limiter}
\lambda=\begin{cases}
\frac{2}{3+\sqrt{9+12\mathcal{R}^{2}}}&\quad\quad0\le \mathcal{R}\le\frac{3}{2}	\\
\frac{1}{1+\mathcal{R}+\sqrt{1+2\mathcal{R}}}&\quad\quad\frac{3}{2}<\mathcal{R}<\infty
\end{cases}
\end{equation}
where
\begin{equation}\label{eqn:reducer}
	\mathcal{R}=\frac{|\nabla\E|}{\kr\rho\E}.
\end{equation}
Other flux limiters are also available \citep{levermore1981,kley1989}.

The Rosseland mean opacity and Planck mean opacity are defined as
\begin{eqnarray}
    \frac{1}{\kr}&=&\frac{\int I_{\nu}/\kappa_{\nu}d\nu}{\int I_{\nu}d\nu},     \\
    \kp&=&\frac{\int\kappa_{\nu}I_{\nu}d\nu}{\int I_{\nu}d\nu}
\end{eqnarray}
where $\kappa_{\nu}$ is the opacity at frequency $\nu$. These mean opacities are often tabulated as a function of radiation temperature $T_{\rm rad}$.
In Equation \ref{eqn:radsource}, the emissivity is by default taken to be,
\begin{equation}
    \eta=\ke\rho c a_{R}\tg^{4},
\end{equation}
where $\ke$ is the emission opacity measured in the co-moving frame. In more realistic cases, $\eta$ can be tabulated \cite[e.g.][]{colombo2019a}. In the test problems used in this work, we assume $\ke=\kp$ for simplicity.

\section{Numerical scheme} \label{sec:method}

We solve the system of general EoS radiation hydrodynamic equations using a two-step operator-splitting approach. In the first step, the pure hydrodynamic equations are resolved using a general EoS Riemann solver (see Section \ref{sec:hydrosolver}). In the second step, we solve the radiation transport equations under the FLD approximation, concurrently accounting for the thermodynamic coupling with the general EoS. This split allows for the use of robust hyperbolic solvers for the fluid dynamics while employing implicit methods for the stiff radiation-matter coupling.

\subsection{Hydrodynamic solvers}

\subsubsection{Basic algorithm for general EoS hydrodynamics}\label{sec:hydrosolver}

Firstly, we solve the hydrodynamic equations and a realistic EoS with a general HLLC Riemann solver \citep{chen2019}, or a classic HLLC Riemann solver in the case of $\gamma$-law gas \citep{batten1997}. In this section, we assume a uniform grid in Cartesian coordinates and focus on describing the necessary steps to solve the equations in 2D by the MUSCL scheme \citep{vanleer1974} which has a second-order spatial and temporal accuracy.
A novel algorithm for spherical polar coordinates is described in Section \ref{sec:angular}.

The hydrodynamic equations are expressed in compact conservation form, temporarily omitting the energy source terms $S_{\text{g}}$ and $G^{0}$:
\begin{equation}\label{eqn:conservhydro}
    \pdv{U}{t}+\pdv{F}{x}+\pdv{G}{y}=S,
\end{equation}
where, $U,\ F,\ G$, and $S$ are the conserved quantities, fluxes in the $x$ and $y$ directions, and the source term due to any external forces, respectively, i.e.,
\begin{equation}\label{eqn:expand}
\begin{split}
U&=\begin{bmatrix}
\rho	\\
\rho v_x	\\
\rho v_y    \\
E
\end{bmatrix},
F=\begin{bmatrix}
\rho v_x	\\
\rho v_{x}^{2}+p	\\
\rho v_{x}v_{y}     \\
v(E+p)
\end{bmatrix},
G=\begin{bmatrix}
\rho v_y	        \\
\rho v_{x}v_{y}	    \\
\rho v_{y}^{2}+p    \\
v(E+p)
\end{bmatrix},   \\
S&=\begin{bmatrix}
0	        \\
\rho f_{x}	    \\
\rho f_{y}    \\
\rho(f_x v_x+f_y v_y)
\end{bmatrix}.
\end{split}
\end{equation}
The primitive quantities in cell $(i,j)$ are denoted by the vector $W_{i,j}^{n} = [\rho, v_x, v_y, p]^{T}$. Our primary strategy is to solve Equation \ref{eqn:conservhydro} in a directionally unsplit manner. Specifically, we treat the hyperbolic flux divergence and the source terms as:
\begin{align}
\pdv{U}{t} &= -\pdv{F}{x} - \pdv{G}{y}, \label{eqn:hydrosplit1} \\
\pdv{U}{t} &= S. \label{eqn:hydrosplit2}
\end{align}

To advance the system from time step $n$ to $n+1$, we employ the following second-order predictor-corrector (midpoint) procedure:

\begin{enumerate}
    \item We perform piecewise linear reconstruction of the primitive quantities $W_{i,j}^{n}$ at all cell interfaces using a slope limiter (SL) \citep{zou2021}. This yields the interface states $W_{i-\frac{1}{2},j,\text{R}}^{n}$ and $W_{i+\frac{1}{2},j,\text{L}}^{n}$, where L and R denote the left and right sides of the interface, respectively. A similar procedure is applied along the $y$-axis. The limited difference $\delta W_{i,j}^{n}$ along the $x$-direction is calculated as:
    \begin{eqnarray}
        W_{i-\half,j,\text{R}}^{n}&=&W_{i,j}-\delta W_{i,j}^{n}/2,     \\
        W_{i+\half,j,\text{L}}^{n}&=&W_{i,j}+\delta W_{i,j}^{n}/2,     \\
        \delta W_{i,j}^{n}&=&\text{SL}(\bar{\delta} W_{i,j,L}^{n},\bar{\delta} W_{i,j,C}^{n},\bar{\delta} W_{i,j,R}^{n}),   \label{eqn:sl}
    \end{eqnarray}
    where $\delta W_{i,j}^{n}$ denotes the limited difference and is calculated from the left, right, and central differences,
    \begin{eqnarray}
        \bar{\delta}W_{i,j,\text{L}}^{n}&=&W_{i,j}^{n}-W_{i-1,j}^{n}   \\
        \bar{\delta}W_{i,j,\text{R}}^{n}&=&W_{i+1,j}^{n}-W_{i,j}^{n}   \\
        \bar{\delta}W_{i,j,\text{C}}^{n}&=&(W_{i+1,j}^{n}-W_{i-1,j}^{n})/2
    \end{eqnarray}
    To ensure numerical robustness---particularly in the presence of complex EoS and curvilinear coordinates---{\tt Guangqi} employs the minmod limiter as its default.
    \item Solve the general EoS Riemann problems with an HLLC Riemann solver \citep{chen2019} at all interfaces given $W_{i\pm\half,j,\text{L/R}}^{n}$ and $W_{i,j\pm\half,\text{L/R}}^{n}$ to obtain the fluxes $F_{i\pm\half,j}^{n}$ and $G_{i,j\pm\half}^{n}$ at the $x$ and $y$ interfaces. In the mean time, save the wave speeds of the Riemann problems at all interfaces.
    \item Estimate the hydrodynamic timestep $\Delta t$ based on the Courant-Friedrichs-Lewy (CFL) condition \citep{courant1928},
    \begin{equation}
        \Delta t=\text{CFL}\cdot\min\bigg\{\min\bigg(\frac{\Delta x}{v_{x}}\bigg),\min\bigg(\frac{\Delta y}{v_{y}}\bigg)\bigg\},
    \end{equation}
    where $\Delta x$ and $\Delta y$ are the cell size in $x$ and $y$ directions, and $v_x$ and $v_y$ are the wave speeds of the Riemann solution at the cells' $x$ and $y$ directions' interfaces. Stability requires the CFL number to be $\le1$ in 1D problems and $\le0.5$ in 2D problems. This criterion also applies to the non-uniform grid.
    \item Integrate the Equation \ref{eqn:hydrosplit1} for half a time-step $\Delta t/2$ by applying flux divergence law
    \begin{equation}
        \begin{split}
            \bar{U}_{i,j}^{n+\half}-U_{i,j}^{n}&=\frac{\Delta t(F_{i-\half,j}^{n}-F_{i+\half,j}^{n})}{2\Delta x}\\&+\frac{\Delta t(G_{i,j-\half}^{n}-G_{i,j+\half}^{n})}{2\Delta y}.
        \end{split}
    \end{equation}
    Integrate the Equation \ref{eqn:hydrosplit2} for half a time-step $\Delta t/2$
    \begin{equation}\label{eqn:source1}
        {U}_{i,j}^{n+\half}-\bar{U}_{i,j}^{n+\half}=\frac{\Delta t S_{i,j}^{n}}{2}.
    \end{equation}
    \item Convert $U_{i,j}^{n+\half}$ to $W_{i,j}^{n+\half}$, reconstruct $W_{i,j}^{n+\half}$ with the same slope limiter to obtain $W_{i\pm\half,j,\text{L/R}}^{n+\half}$ and $W_{i,j\pm\half,\text{L/R}}^{n+\half}$.
    \item Solve the Riemann problems at all interfaces to obtain $F_{i\pm\half,j}^{n+\half}$ and $G_{i,j\pm\half}^{n+\half}$.
    \item Apply flux divergence law to calculate $\bar{U}^{n+1}$,
    \begin{equation}
        \begin{split}
            \bar{U}_{i,j}^{n+1}-U_{i,j}^{n}&=\frac{\Delta t(F_{i-\half,j}^{n+\half}-F_{i+\half,j}^{n+\half})}{\Delta x}\\&+\frac{\Delta t(G_{i,j-\half}^{n+\half}-G_{i,j+\half}^{n+\half})}{\Delta y},
        \end{split}
    \end{equation}
    and integrate the Equation \ref{eqn:hydrosplit2} for $\Delta t$ from the state of time step $n$.
    \begin{equation}\label{eqn:source2}
        {U}_{i,j}^{n+1}-\bar{U}_{i,j}^{n+1}=\Delta t S_{i,j}^{n+\half}.
    \end{equation}
    \item Convert ${U}_{i,j}^{n+1}$ to $W_{i,j}^{n+1}$.
\end{enumerate}
 
We have not included boundary communications here. They will be discussed in Section \ref{sec:hypercomm}, where additional communications associated with static mesh refinement (SMR)/AMR will also be addressed.

\subsubsection{Extension to spherical-polar coordinates: the PSAMA
}\label{sec:angular}

Astrophysical simulations in spherical-polar coordinates frequently involve systems dominated by central gravity, where the precise conservation of angular momentum is paramount.
We introduce a novel numerical scheme that conserves angular momentum to machine precision while maintaining superior numerical stability compared to conventional methods.
The derivation here assumes azimuthal symmetry (commensurate with the 2D nature of {\tt Guangqi}), but
generalization to 3D is straightforward. We first write down the hydrodynamic equations in spherical polar coordinates,
\begin{eqnarray}
	\pdv{\rho}{t}+\frac{1}{r^{2}}\pdv{(r^{2}\rho v_{r})}{r}+\frac{1}{r\sin\theta}\pdv{(\sin\theta\rho v_{\theta})}{\theta}=0,&\quad\enspace	\label{eqn:spher1}\\
    \pdv{(\rho v_{r})}{t}+\frac{1}{r^2}\pdv{[r^{2}(\rho v^{2}_{r}+p)]}{r}+\frac{1}{r\sin\theta}\pdv{(\sin\theta\rho v_{r}v_{\theta})}{\theta}=&\nonumber\\
    \rho g_{r}+\frac{2p}{r}+\frac{\rho(v^{2}_{\theta}+v^{2}_{\phi})}{r},	\label{eqn:spher2}\\
	\pdv{(\rho v_{\theta})}{t}+\frac{1}{r^2}\pdv{(r^{2}\rho v_{r}v_{\theta})}{r}+\frac{1}{r\sin\theta}\pdv{[\sin\theta(\rho v^{2}_{\theta}+p)]}{\theta}=&\nonumber\\
    \frac{\cot\theta(\rho v^{2}_{\phi}+p)-\rho  v_{r}v_{\theta}}{r},	\label{eqn:spher3}\\
	\pdv{(\rho v_{\phi})}{t}+\frac{1}{r^2}\pdv{(r^{2}\rho v_{r}v_{\phi})}{r}+\frac{1}{r\sin\theta}\pdv{(\sin\theta\rho v_{\theta}v_{\phi})}{\theta}=&\nonumber\\
    -\frac{\rho v_{r}v_{\phi}}{r}-\frac{\cot\theta\rho v_{\theta}v_{\phi}}{r},	\label{eqn:spher4}\\
	\pdv{E}{t}+\frac{1}{r^{2}}\pdv{[r^{2}(E+p) v_{r}]}{r}+\frac{1}{r\sin\theta}\pdv{[\sin\theta (E+p) v_{\theta}]}{\theta}=&\nonumber\\
    \rho g_{r}v_{r}.	\label{eqn:spher5}
\end{eqnarray}
The left-hand side (LHS) of Equations \ref{eqn:spher1}--\ref{eqn:spher5} represents the system in linear-momentum conservative form, which can be resolved directly using a standard Riemann solver at cell interfaces. The complete evolution is achieved by integrating the source terms on the right-hand side (RHS) following the second-order predictor-corrector steps described in Equations \ref{eqn:source1} and \ref{eqn:source2}. However, due to finite spatial resolution, the non-zero source terms---particularly those in the $\phi$-momentum equation---generally lead to a violation of the underlying conservation laws. Specifically, the conservation of linear momentum in the $\phi$-direction and the resulting angular momentum along the polar axis are not guaranteed due to the geometric source terms in Equation \ref{eqn:spher4}.

A common remedy is to post-process the linear momentum flux after the Riemann solver stage to reflect angular momentum conservation in curvilinear coordinates \citep{mignone2007,skinner2010,ju2016,ju2017}. Nevertheless, this modification does not recover the true solution of an angular-momentum-based Riemann problem. Furthermore, in the modification, the conserved quantities is changed from the linear momentum in the Riemann problem solution to the angular momentum that is used in the flux divergence law, which introduces an inconsistency (see Appendix \ref{app:riemannproblems} for further discussions). To overcome the tension between the angular momentum conservation and consistent energy evolution when using a Riemann solver, we propose a new scheme, which we term ``passive scalar angular momentum algorithm (PSAMA)" as follows. 

We first rewrite Equation \ref{eqn:spher4} in the angular momentum conservation form,
\begin{eqnarray}
    \pdv{(\rho l)}{t}+\frac{1}{r^2}\pdv{(r^{2}\rho v_{r}l)}{r}+\frac{1}{r\sin\theta}\pdv{(\rho v_{\theta}l\sin\theta)}{\theta}&=&\nonumber\\
    \pdv{(\rho l)}{t}+\nabla\cdot(\rho\vec{v}l)&=&0\label{eqn:spher4am}
\end{eqnarray}
where $l=r\sin\theta v_{\phi}=(r\sin\theta)^{2}\Omega$ is the specific angular momentum along the polar axis and $\Omega$ is the angular frequency. Equation \ref{eqn:spher4am} indicates that the angular momentum is a passive scalar. Then, we split the total energy into two parts in the following way,
\begin{equation}\label{eqn:energysplit}    E=\underbrace{\rho\bigg(\epsilon_{g}+\frac{v_{r}^{2}+v_{\theta}^{2}}{2}\bigg)}_{E'}+\underbrace{\frac{\rho l^{2}}{2(r\sin\theta)^{2}}}_{e_{k\phi}}
\end{equation}
where $e_{k\phi}$ is the kinetic energy in the $\phi$ direction and $E'$ is the sum of the internal energy and the kinetic energy in $r$ and $\theta$ direction. We adapt Equation \ref{eqn:spher5} correspondingly,
\begin{eqnarray}
    \pdv{E'}{t}+\frac{1}{r^{2}}\pdv{[r^{2}(E'+p) v_{r}]}{r}+\frac{1}{r\sin\theta}\pdv{[\sin\theta(E'+p)v_{\theta}]}{\theta}\nonumber&\\
    +\pdv{e_{k\phi}}{t}+\nabla\cdot({e_{k\phi}\vec{v}})=\rho g_{r}v_{r}.\quad\quad&\label{eqn:esplit}
\end{eqnarray}
The key is to express ${\partial e_{k\phi}}/{\partial t}+\nabla\cdot({e_{k\phi}\vec{v}})$ in terms of known variables. We put the derivation in Appendix \ref{app:derivation}. The result is,
\begin{equation}
    \pdv{e_{k\phi}}{t}+\nabla\cdot{(e_{k\phi}\vec{v})}=-\rho\vec{f}_{\Omega}\cdot\vec{v},\label{eqn:ekphi4}
\end{equation}
where $\vec{f}_{\Omega}=r\Omega^{2}\sin\theta[\sin\theta,\cos\theta,0]^{T}$. Substitute Equation \ref{eqn:ekphi4} into \ref{eqn:esplit}, and move the source term to the RHS, we obtain the energy transport equation,
\begin{equation}
    \pdv{E'}{t}+\nabla\cdot[(E'+p)\vec{v}]=\rho g_{r}v_{r}+\rho\vec{f}_{\Omega}\cdot\vec{v}.\label{eqn:ecorrect}
\end{equation}

To summarize succinctly, we outline the main steps for proceeding a half timestep, which corresponds to steps 1 to 5 in Section \ref{sec:hydrosolver}.
\begin{enumerate}
    \item Reconstruct the primitive quantities. Use $\Omega$ to do the reconstruction of $l$. 
    \item Solve the Riemann problems of $[\rho, \rho v_{r}, \rho v_{\theta}, E']$ in the $r$ and $\theta$ directions, i.e., Equation \ref{eqn:spher1}, \ref{eqn:spher2}, \ref{eqn:spher3}, and \ref{eqn:ecorrect} in the same manner as in the Cartesian coordinate.
    \item Estimate the timestep $\Delta t$.
    \item Apply the flux divergence law to the conserved quantities $[\rho, \rho v_{r}, \rho v_{\theta}, E']$ and the angular momentum $\rho l$. Add the second order source terms as described in Appendix \ref{app:sourceterms} according to the operator unsplit manner introduced in Section \ref{sec:hydrosolver}.
    \item Obtain the half timestep state $[\rho, \rho v_{r}, \rho v_{\theta}, E']$ and $\Omega$ by converting the conserved quantities to the primitive quantities.
\end{enumerate}
Step 6 to 8 follow the same procedures.

PSAMA ensures the precise conservation of angular momentum in spherical-polar coordinates while maintaining consistency between the linear-momentum-based Riemann solver and the flux divergence law. This consistency is achieved at the expense of an additional source term, $\rho\vec{f}_{\Omega}\cdot\vec{v}$, on the right-hand side (RHS) of Equation \ref{eqn:ecorrect}, which formally compromises exact energy conservation. However, in the presence of gravitational source terms and implicit radiation transport, absolute energy conservation is already numerically constrained by the convergence limits of iterative solvers. We therefore advocate for PSAMA as a more robust and internally consistent algorithm for rotating flows. Its effectiveness is demonstrated through a suite of numerical benchmarks in Section \ref{sec:angularmomentumtest}.

\subsection{FLD Radiation transport and implicit method}\label{sec:rad}

The second main step is to solve the equation of radiation transport with the FLD approximation, coupled with the equation of gas thermodynamics,
\begin{eqnarray}
	\pdv{\E}{t}+\nabla\cdot(v\E)-\nabla\cdot(D\pdv{\E}{x})&=&G^{0}	\label{eqn:radtran1}\\
	C_{V}\pdv{\tg}{t}&=&-G^{0}+S_{\text{g}}	\label{eqn:radtran2}
\end{eqnarray}
where
\begin{equation}
	C_{V}=\pdv{e_{\text{g}}(\rho,\tg)}{\tg}
\end{equation}
is the heat capacity per unit volume at $(\rho,\tg)$.

The diffusion term in Equation \ref{eqn:radtran1} imposes a stringent constraint on the simulation timestep if treated explicitly, as the stability criterion $\delta t < \min(\Delta x^{2}/(4D))$ must be satisfied. According to Fick's law, $D \sim v L_{d}$, where $v$ is the characteristic speed and $L_{d}$ is the diffusion length scale. In optically thin regimes, the diffusion coefficient scales as $D \sim c \lambda_{\rm mfp}$, where $\lambda_{\rm mfp}$ is the photon mean free path. Consequently, the required explicit stability limit $\delta t$ becomes orders of magnitude smaller than the hydrodynamic timestep $\Delta t$ governed by the CFL condition (i.e., $\delta t \ll \Delta t$). To circumvent this severe constraint, a fully implicit scheme is required to solve Equations \ref{eqn:radtran1} and \ref{eqn:radtran2}. We therefore adopt a standard backward Euler discretization for the radiation transport.

\subsubsection{Uniform Cartesian grid}\label{sec:uniformgrid}

We start with the algorithm to solve the problem in a uniform Cartesian grid with $N_{1}\times N_{2}$ cells. The algorithm for SMR/AMR grid will be discussed in Section \ref{sec:implicitamr}. We discretize equations \ref{eqn:radtran1} and \ref{eqn:radtran2} using the backward Euler method,
\begin{multline}\label{eqn:linearsys1}
    \phi_{1}E_{i,j-1}^{n+1}+\phi_{2}T_{i,j-1}^{n+1}+\phi_{3}E_{i-1,j}^{n+1}+\phi_{4}T_{i-1,j}^{n+1}\\+\phi_{5}E_{i,j}^{n+1}+\phi_{6}T_{i,j}^{n+1}+\phi_{7}E_{i+1,j}^{n+1}+\phi_{8}T_{i+1,j}^{n+1}\\+\phi_{9}E_{i,j+1}^{n+1}+\phi_{10}T_{i,j+1}^{n+1}=\phi_{11}\ ,
\end{multline}
\begin{multline}\label{eqn:linearsys2}
    \psi_{1}E_{i,j-1}^{n+1}+\psi_{2}T_{i,j-1}^{n+1}+\psi_{3}E_{i-1,j}^{n+1}+\psi_{4}T_{i-1,j}^{n+1}\\+\psi_{5}E_{i,j}^{n+1}+\psi_{6}T_{i,j}^{n+1}+\psi_{7}E_{i+1,j}^{n+1}+\psi_{8}T_{i+1,j}^{n+1}\\+\psi_{9}E_{i,j+1}^{n+1}+\psi_{10}T_{i,j+1}^{n+1}=\psi_{11}\ ,
\end{multline}
where we have omitted the subscripts $g$ in $\tg$ and $r$ in $\E$ for brevity. Equations \ref{eqn:linearsys1} and \ref{eqn:linearsys2} consist of a large sparse linear system (linear system, hereafter) $\mathbb{A}x=\mathbf{b}$, where $\mathbb{A}$ is the matrix of the linear system, $x=[E_{i,j},T_{i,j}]^{T}$ that spans over the entire grid ($2N_{1}N_{2}$ variables to be solved),
and $\mathbf{b}$ represents the right-hand side of the linear system. 

To assemble the linear system, we need to calculate all the $\phi$ and $\psi$ coefficients. First, we approximate the non-linear term $\tg^{4}$ in $G^{0}$ with Taylor's expansion to the first order,
\begin{equation}   
    (\tg^{n+1})^{4}\approx4(\tg^{n})^{3}\tg^{n+1}-3(\tg^{n})^{4}.
\end{equation}
Then, we explicitly calculate the time averaged advection term. Specifically,
\begin{equation}
	\nabla\cdot(v\E)=\nabla\cdot(\rho v\frac{\E}{\rho})=-S_{\text{adv}}.
\end{equation}
We observe that the mass flux $\rho v$ at the cell interfaces corresponds directly to the solution of the mass conservation equation in the Riemann problem, which we denote as $f_{i\pm1/2,j}$ and $f_{i,j\pm1/2}$. Consequently, by assuming that the specific radiation energy, $e_r = E_r/\rho$, remains constant within each cell, the advection term for the $(i,j)$ cell can be discretized as follows:
\begin{equation}\label{eqn:adv}
    \begin{split}
        S_{\text{adv}}&=g_{i-\half,j}f_{i-\half,j}e_{\rm{r},i-\half,j}-g_{i+\half,j}f_{i+\half,j}e_{\rm{r},i+\half,j}
        \\&+g_{i,j-\half}f_{i,j-\half}e_{\rm{r},i,j-\half}-g_{i,j+\half}f_{i,j+\half}e_{\rm{r},i,j+\half},
    \end{split}
\end{equation}
where
\begin{equation}
    g_{i\pm\half,j\pm\half}=dS_{i\pm\half,j\pm\half}/dV_{i,j}, \label{eqn:fldg}
\end{equation}
is the geometric factor, and $dS_{i\pm\half,j\pm\half}$ and $dV_{i,j}$ are the corresponding cell's bounding interface area and volume. The specific radiation energy $e_{r,i\pm\half,j\pm\half}$ can be determined by the upwind condition.

On a uniform grid, the diffusion coefficient $D$ is calculated at all interfaces between neighboring cells. Taking the interfaces perpendicular to the $x$-axis as an example, the coefficient is discretized as:
\begin{equation}\label{eqn:diffcoeff}
        D_{i+\half}=\frac{-c\lambda(\mathcal{R})}{(\kr\rho)_{i+\half}},
\end{equation}
where $(\kr\rho)_{i+\half}=\sigma_{i+\half}$ is an interface value and should be interpolated from the cell center value. To determine the dimensionless gradient $\mathcal{R}$, which is required for the flux limiter $\lambda(\mathcal{R})$, we utilize the spatial average of the radiation energy density from the adjacent cells to estimate the interface value $E_{r,i+\half}$.

\begin{figure}
	\centering
	\includegraphics[width=\columnwidth]{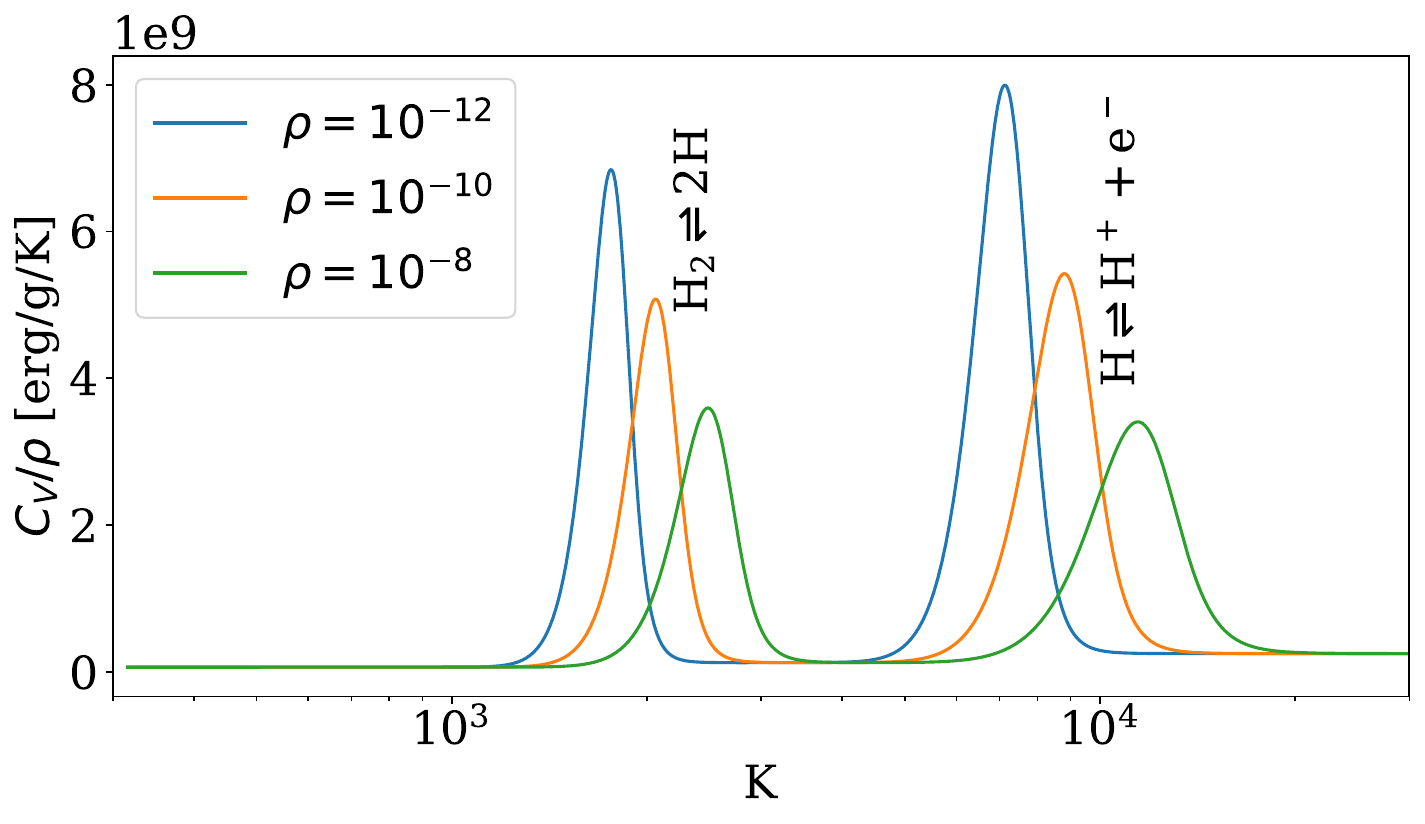}
	\caption{The specific heat capacity $C_{V}/\rho$ of hydrogen gas as a function of temperature.
    The yellow and purple bands indicate the dissociation and ionization region, respectively, where the specific heat capacity shows rapid variations.}
	\label{fig:cv}
\end{figure}

The non-zero elements of the linear system are:
\begin{eqnarray}
    \phi_{1}&=&-\alpha_{i,j-\half} g_{i,j-\half}D_{i,j-\half}^{n},	\label{eqn:phi1}\\
    \phi_{3}&=&-\alpha_{i-\half,j} g_{i-\half,j}D_{i-\half,j}^{n},	\\
    \phi_{5}&=&1-\phi_{1}-\phi_{3}-\phi_{7}-\phi_{9}+\beta_{P}^{n}	\\
    \phi_{6}&=&-4\beta_{e}^{n}a_{R}\big(T_{i,j}^{n}\big)^{3},	\\
	\phi_{7}&=&-\alpha_{i+\half,j} g_{i+\half,j}D_{i+\half,j}^{n},	\\
	\phi_{9}&=&-\alpha_{i,j+\half} g_{i,j+\half}D_{i,j+\half}^{n},	\\	
    \phi_{11}&=&E_{r,i,j}^{n}-3\beta_{e} a_{R}\big(T_{i,j}^{n}\big)^{4}+(S_{\text{adv}}+S_{\text{g}})\delta t    \\
    \psi_{5}&=&-\beta_{P}^{n}	\\
    \psi_{6}&=&C_{V}+4\beta_{e} a_{R}(T_{i,j}^{n})^{3}	\\
    \psi_{11}&=&C_{V}T_{i,j}+3\beta_{e} a_{R}(T_{i,j}^{n})^{4}   \label{eqn:psi7}
\end{eqnarray}
where $\beta_{P}=\kp\rho c\delta t$, $\beta_{e}=\ke\rho c\delta t$, and $\alpha_{i\pm\half,j\pm\half}$ is related to the spacing between the neighboring cells, for example,
\begin{equation}\label{eqn:alpha}
    \alpha_{i+\half,j}=\delta t/(x_{i+1,j}-x_{i,j}),
\end{equation}
where $x_{i,j}$ is volume center. Here $\delta t\le\Delta t$ is the timestep of the radiation transport step. When $\delta t=\Delta t$, we integrate the radiation transport equation in a single time step. However, the gas and radiation energy coupling strength may be very strong and the $\tg$ may be drastically different from the $\trad$ at shocks, making the local thermodynamics a stiff problem. In addition, $C_{V}$ of a realistic gas may vary rapidly with temperature during phase transitions and can be another stiff term. As an example, Figure \ref{fig:cv} shows $C_{V}/\rho$ v.s. $\tg$ of pure hydrogen gases at various densities.

The most straightforward recipe to handle the stiff problem is to reduce the integration timestep. In principle, we can split the integration of the radiation transport into sub-steps that is determined by
\begin{eqnarray}\label{eqn:substepglobal}
    \delta t&=&\min(\Delta t_{\text{remain}},\delta t_{\text{rad},i,j})    \\
    \delta t_{\text{rad},i,j}&=&\bar{\delta}\tg\rho C_{V}/|G^{0}|
\end{eqnarray}
where $\Delta t_{\rm remain}$ is the remaining time of this hydro time step, $\delta t_{\rm{rad},i,j}$ is the time step limited by the radiation-matter coupling strength $G^{0}$ at $(i,j)$ cell, and $\bar{\delta}<0.1$ is the coefficient that controls the maximum temperature change between each sub-step.
In practice, we can use a prescribed geometric series to determine the sub-timesteps,
\begin{eqnarray}
    \delta t_{i_{\text{sub}}}&=&\delta t_{0}q_{t}^{i_{\text{sub}}-1},\quad\quad i_{\text{sub}}\in[1,n_{\text{sub}}] \label{eqn:subtimesteps}\\
    \delta t_{0}&=&\frac{\Delta t(1-q_{t})}{1-q_{t}^{n_{\text{sub}}}},
\end{eqnarray}
where $1<q_{t}<2$ is the geometric factor and $n_{\text{sub}}$ is total number of sub-cycles. The choice of $q_{t}$ and $n_{\text{sub}}$ are very problem dependent and should be explored by the users. We present several tests in Section \ref{sec:radcouplinghydrogen}.

The rank of the linear system (Equation \ref{eqn:radtran1} and \ref{eqn:radtran2}) is $2N_{1}N_{2}$, and the number of non-zero elements are small compared to the number of zero elements. To solve the linear system efficiently, we employ the external package {\tt Petsc}\footnote{\href{https://petsc.org/release/}{{\tt Petsc} webpage}}, and use the GMRES iteration method \citep{saad1986}. We have experimented with some classic preconditioners including the block Jacobi, incomplete LU, and LU preconditioners, and found that additive Schwarz preconditioner (PCASM\footnote{\href{https://petsc.org/release/manualpages/PC/PCASM/}{{\tt Petsc} PCASM webpage}}) is in general a good choice for the block structure computational domain. As an iterative solver, {\tt Petsc} solves the linear system problem $\mathbb{A}\mathbf{x}=\mathbf{b}$ approximately. We set the convergence criterion by the relative error $\er$ in the residual,
\begin{equation}\label{eqn:linearsystem}
    \|\mathbf{b}-\mathbb{A}\mathbf{x}_{k}\|<\er\|\mathbf{b}\|,
\end{equation}
where $\|\cdot\|$ takes the 2-norm and $k$ in the subscript is the iteration number.

Finally, we can calculate $\F$ at interfaces as,
\begin{equation}
    \vec{F}_{r,i+\frac{1}{2},j}=D^{n}_{i+\frac{1}{2},j}\sum_{i_{\text{sub}}}\frac{\alpha_{i+\half,j}^{i_{\text{sub}}}(E^{n,i_{\text{sub}}}_{r,i+1,j}-E^{n,i_{\text{sub}}}_{r,i,j})}{\Delta t}
\end{equation}
where $i_{\text{sub}}$ is the sub-cycle index.

We summarize the calculation steps in the radiation subsystem.
\begin{enumerate}
    \item Calculate $S_{\text{adv}}$ according to Equation \ref{eqn:adv} of all cells. Construct the series of $\delta t_{i_{\text{sub}}}$ according to Equation \ref{eqn:subtimesteps}.
    \item Calculate $g$, $D$, and $\alpha$ and at all interfaces according to Equation \ref{eqn:fldg}, \ref{eqn:diffcoeff}, and \ref{eqn:alpha}, respectively.
    \item Calculate $C_V$ of all cells.
    \item Calculate $\phi$ and $\psi$ with Equation \ref{eqn:phi1}-\ref{eqn:psi7}.
    \item Assemble $\mathbb{A}$ and $\mathbf{b}$ of the linear system.
    \item Solve the linear system problem with an iterative solver.
    \item When $i_{\text{sub}}<n_{\text{sub}}$, update $\alpha$ and go to step 3.
    \item Calculate $\vec{F}_{r}$ if needed.
\end{enumerate}

\subsubsection{Implicit algorithm with SMR/AMR}\label{sec:implicitamr}

\begin{figure}
    \centering
    \includegraphics[width=\columnwidth]{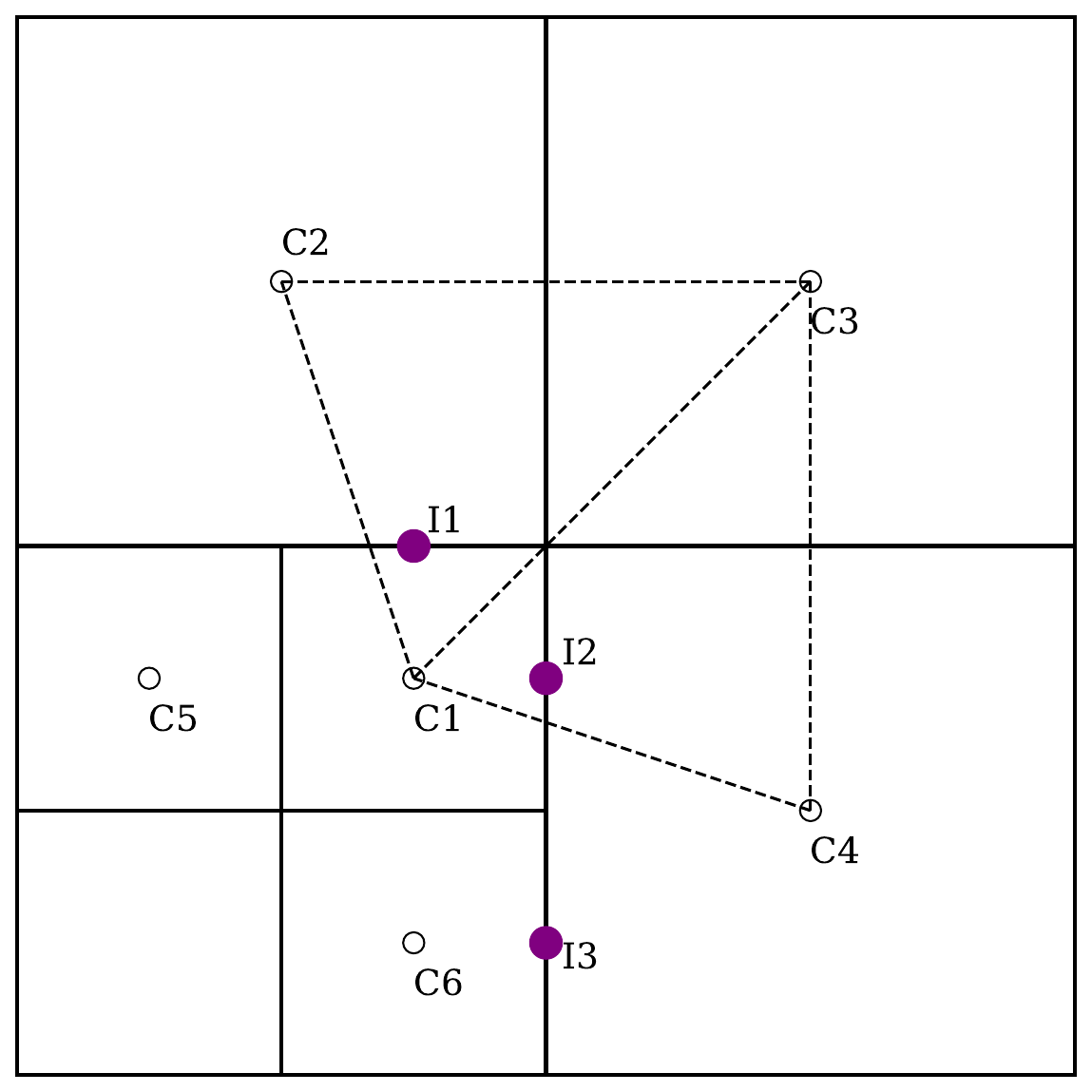}
    \caption{A mesh refinement structure showing refined cells in the bottom-left next to the unrefined cells. Values of interfaces (purple dots labeled I) and cell centers (black hollow circles) when there is a level difference.}
    \label{fig:interp}
\end{figure}

The major difference between a uniform grid and a grid with SMR/AMR in the radiation transport step is the topological relation between different levels. Thus, it leads to a more complex calculation of the diffusion coefficient $D$. As mentioned in Section \ref{sec:uniformgrid}, we need to calculate $\E$, $\nabla\E$, and $\kr\rho$ at all the interfaces. In a grid with mesh refinement (Figure \ref{fig:interp}), the interface value is related to more neighbor cells.

Figure \ref{fig:interp} shows refined cells in the bottom-left next to the coarser cells, where I1, I2, and I3 are at the interfaces with the finer cells. We can linearly interpolate $\E$ and $\kr\rho$ at C1, C2, and C3 to get $\E$, $\nabla\E$, and $\kr\rho$ at I1, the same logic follows for I2 and I3. Then, the $D$ of the finer cells at the level jumps can be calculated from the interpolated variables. However, the value of $D$ for the coarser cells at the level jumps is more complicated.

In the AMR implementations described by \citet{commercon2011,commercon2014}, a single $D$ of the coarse cells at the level jumps is calculated with interpolated variables at the interfaces so that all cells have the same number of neighbors. The linear system is thus more regular with this topological configuration. However, there is a potential numerical inconsistency between the total fluxes crossing the fine cells and the flux crossing the coarse cell if they are using different $D$.

Alternatively, one can adopt the ``deferred synchronization" algorithm that solves the partitioned matrix across different refinement levels in a staggered manner \citep{howell2003,zhang2011,ramsey2015}. The load balancing suffers from this additional splitting and radiation transport is also splitted, i.e., the refined zone does not know the physical condition (optically thick/thin) of the coarse zone when it is being updated, vice versa.

Here, we let the coarse cells to communicate directly with the fine cells and solve the multilevel implicit radiation transport problem in a single matrix.
Specifically, we send $D$ at I2 and I3 to the coarse cell at C4, and C4 will have 5 neighboring cells (assuming other neighbors are the same level as C4). Correspondingly, the linear system of C4 has one additional non-zeros terms, and changes to
\begin{multline}\label{eqn:linearsys1_amr}
    \phi_{1}E_{i,j-1}^{n}+\phi_{3,C1}E_{i-1,j,C1}^{n}+\phi_{3,C6}E_{i-1,j,C6}^{n}+\\+\phi_{5}E_{i,j}^{n}+\phi_{6}T_{i,j}^{n}+\phi_{7}E_{i+1,j}^{n}+\phi_{9}E_{i,j+1}^{n}=\phi_{11}.
\end{multline}
The presence of $\phi_{3,C1}E_{i-1,j,C1}^{n}$ and $\phi_{3,C6}E_{i-1,j,C6}^{n}$ enables the direct communication between C1, C4, and C6. Meanwhile, the fluxes between the fine and coarse cells are now more consistent.

\section{Domain decomposition and communications}\label{sec:framework}

As an SMR/AMR code employing MPI parallelization, {\tt Guangqi} manages complex data synchronization across multiple refinement levels and processor domains. Our 2D AMR framework utilizes a block-structured quad-tree data structure, a design philosophy shared with established codes such as {\tt FLASH} \citep{fryxell2000}, {\tt AMRVAC} \citep{keppens2003}, and {\tt Athena++} \citep{stone2020}.

The criteria for mesh refinement are highly problem-dependent; for a comprehensive list of representative criteria, we refer the reader to \citet{bryan2014}. Currently, to ensure simplicity and maximize parallel efficiency, {\tt Guangqi} integrates all cells across different levels using a global time step $\Delta t$. While this approach simplifies the global synchronization of the monolithic radiation matrix, it presents challenges in multilevel configurations. Specifically, the small $\Delta t$ required by the finest level effectively reduces the CFL number for the coarser grids, thereby introducing significant numerical diffusion. For simulations that prioritize the accurate advection of large-scale structures on the coarse grid, an adaptive, level-dependent timestep may be preferable \citep{bryan2014,weinberger2020,wibking2022}, which we will consider implementing in the future.

\subsection{Domain decomposition}\label{sec:DDLB}

\begin{figure}
    \centering
    \includegraphics[width=\columnwidth]{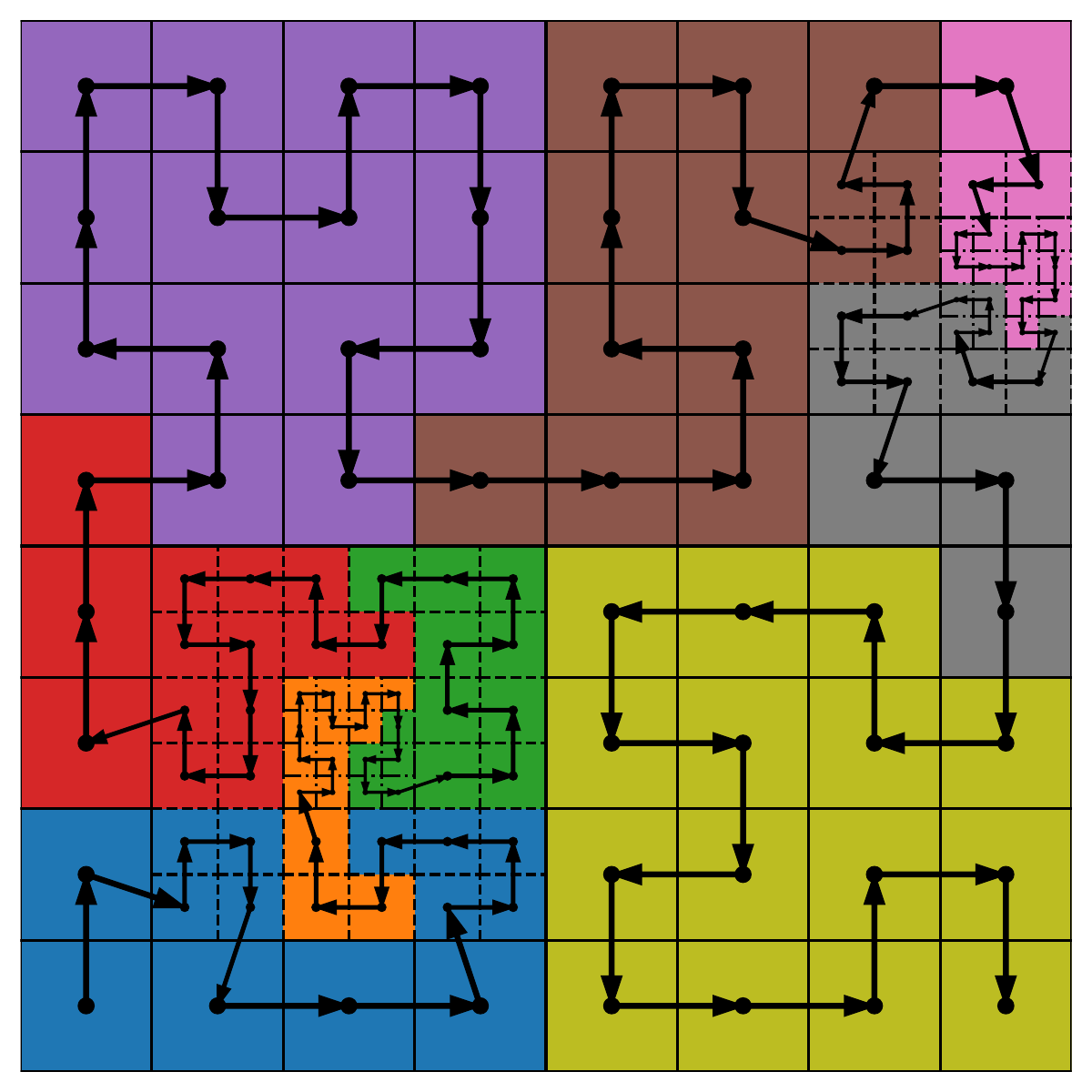}
    \includegraphics[width=\columnwidth]{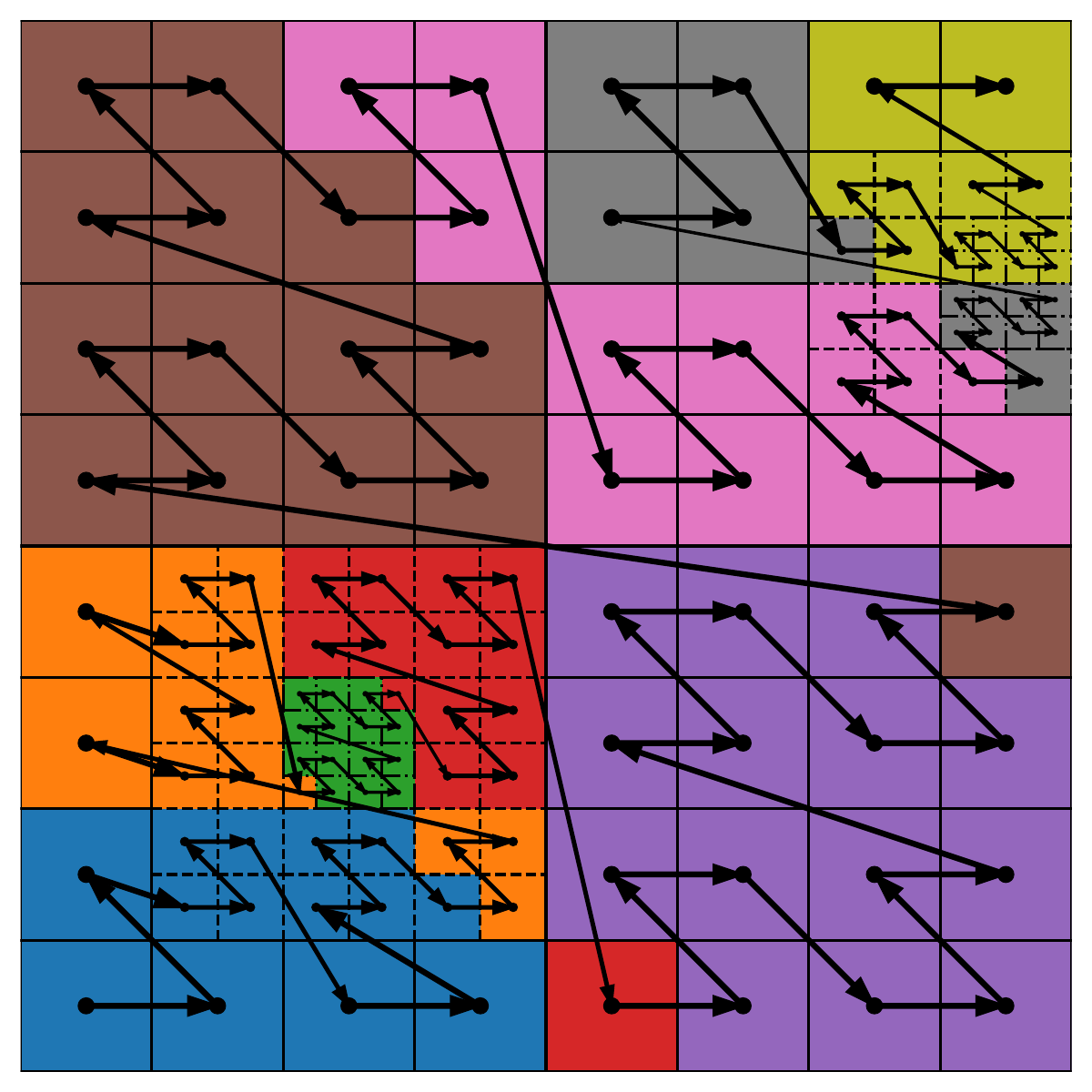}
    \caption{Example of domain decomposition with the same mesh refinement. The computational domain is distributed into 9 sub-domains, represented by different colors. The upper panel shows the domain decomposition of Hilbert curve, and the bottom panel shows the domain decomposition of Z-order curve.}
    \label{fig:domaindecom}
\end{figure}

A 2D/3D block-structured computational domain can be mapped to a 1D list by a space-filling curve. The 1D list can then be subdivided and distributed to different memory/cores in a distributed computing system.
Popular space-filling curves include the Z-order filling curve that has been adopted by {\tt AMRVAC} and {\tt Athena++} or the Hilbert filling curve \footnote{{\href{https://en.wikipedia.org/wiki/Hilbert\_curve}{Hilbert space-filling curve.}}} that is recently implemented in {\tt Kratos} \citep{wang2025} and is also adopted by {\tt Guangqi}. Figure \ref{fig:domaindecom} illustrates the difference between the two space-filling curves on the same computational domain. Each color denotes a sub-domain. We can see that the Z-order curve may generate disconnected sub-domains, increase the surface-to-volume ratio, and require more communication between sub-domains. This is because when a 2D/3D domain is mapped to 1D, the blocks that are close in 2D/3D are not necessarily close in 1D, and vice versa. In principle, reducing the communication workload could improve the parallel performance. However, in our practice, we find that the performance between the two space-filling curves is insignificant. We conjecture that {\tt Guangqi} may not have optimized to the level that this subtle difference can clearly emerge, or the surface-to-volume ratio problem is not a severe bottleneck in 2D problems. Thus, both the Hilbert and Z-order curves are implemented in {\tt Guangqi}.

\subsection{Communications for hydrodynamic variables}\label{sec:hypercomm}

Hydrodynamic solvers benefit from the uniform data structure within a block, but the cells near the blocks boundaries require prolongation, restriction and communication where there is a level difference. We outline the necessary steps to communicate between the coarse and fine blocks' boundary in {\tt Guangqi}.

\begin{enumerate}
    \item Do cell-wise restriction on the fine block's boundary based on the conservation laws,
    \begin{equation}
        U_{i,j}=\frac{\sum_{2i-1,2j-1}^{2i,2j}U'_{i',j'}}{\sum_{2i-1,2j-1}^{2i,2j}V'_{i',j'}},
    \end{equation}
    where $V'$ is the cell volume on the fine grid, $U_{i,j}$ is the conserved variables on the restricted grid. Save the restricted cells for prolongation.
    \item Pass the restricted cells to the guard cells of the coarse block. In the mean time, pass the original cells of the coarse blocks to the guard cells of the fine blocks.
    \item Prolongate the primitive variables on the guard cells with the received data from the coarse block and the previously saved restricted local cells of fine blocks, utilizing a reconstruction method,
    \begin{align}
        W_{2i-1,2j-1}&=&W'_{i,j}-\delta_{x} W'_{i,j}-\delta_{y} W'_{i,j},  \\
        W_{2i-1,2j}&=&W'_{i,j}-\delta_{x} W'_{i,j}+\delta_{y} W'_{i,j},  \\
        W_{2i,2j-1}&=&W'_{i,j}+\delta_{x} W'_{i,j}-\delta_{y} W'_{i,j},  \\
        W_{2i,2j}&=&W'_{i,j}+\delta_{x} W'_{i,j}+\delta_{y} W'_{i,j},
    \end{align}
    where $\delta_{x} W'_{i,j}$ and $\delta_{y} W'_{i,j}$ are the estimated differences between the prolongated variables ($W$) and the original variables ($W'$), utilizing the slope limited primitive variables (see Equation \ref{eqn:sl}) in the $x$ and $y$ coordinates.
\end{enumerate}

In the multilevel communication, the prolongation can be done to the primitive variables with van Leer slope limiter reconstruction without breaking the conservation law, because this prolongation is done on the guard cells of the fine grid, and we only need them to calculate the fluxes on the block boundaries. Notably, in block-wise prolongation in the SMR/AMR\footnote{Sometimes we need to restart simulations and add refinements to specific regions.}, the prolongation must be done to the conservative variables with a less aggressive slope limiter, such as the minmod, to satisfy conservation laws and maintain numerical stability \citep{stone2020}.

After the calculation of the fluxes at the interfaces, we also need to do flux correction on the coarse cells to satisfy the conservation law. The flux correction step in 2D is
\begin{equation}
    F_{\text{coarse},j}=\frac{F_{\text{fine},2j-1}\mathcal{A}_{\text{fine},2j-1}+F_{\text{fine},2j}\mathcal{A}_{\text{fine},2j}}{\mathcal{A}_{\text{fine},2j-1}+\mathcal{A}_{\text{fine},2j}},
\end{equation}
where $F$ is defined in Equation \ref{eqn:expand} and $\mathcal{A}$ is the interface area. Passive scalars follow the same communication pattern as the hydrodynamic variables. The communication between the same level blocks are based on MPI derived datatypes and do not require flux correction.

\subsection{Communications for radiation variables}

As detailed in Section \ref{sec:implicitamr}, the monolithic matrix is assembled by interpolating cell-centered quantities ($\E$ and $\kr\rho$) to determine the interface-centered variables ($\E$, $\kr\rho$, $\nabla\E$, and $D$). During the interpolation stage, the precise spatial coordinates of the neighboring cell centers (e.g., C2, C3, and C6) are required to compute the interface values at I1 and I2.

Consequently, the solver only requires a precomputation of neighboring cell locations based on the underlying grid hierarchy. By explicitly accounting for these interface interactions within the global matrix, the implicit solver maintains flux consistency across refinement levels without requiring standard prolongation or restriction operators。

\section{Numerical tests}\label{sec:tests}

In this section, we test the capability of our code in solving the general EoS hydrodynamics, radiation transport, and radiation-hydrodynamics. We also test the functionality of SMR/AMR, and the spherical coordinate system.

\subsection{Radiation-matter coupling (0D) tests}

\subsubsection{Radiation and perfect gas coupling}\label{sec:perfectcoupling}

We start by examining the energy coupling between radiation and a perfect gas. Following \cite{turner2001} and \cite{kolb2013}, we set up a 1D domain with a constant and uniform radiation energy density $\E=10^{12}$ \ergcmc\ and 256 cells in the Cartesian geometry. Therefore, the rank of the linear system is 512. The gas internal energy $e_{\text{g}}$ should evolve according to,
\begin{equation}
    \pdv{e_{\text{g}}}{t}=-\kp\rho c(a_{R}\tg^{4}-\E)  \label{eqn:radmattercoupling}
\end{equation}

\begin{figure}
	\centering
	\includegraphics[width=\columnwidth]{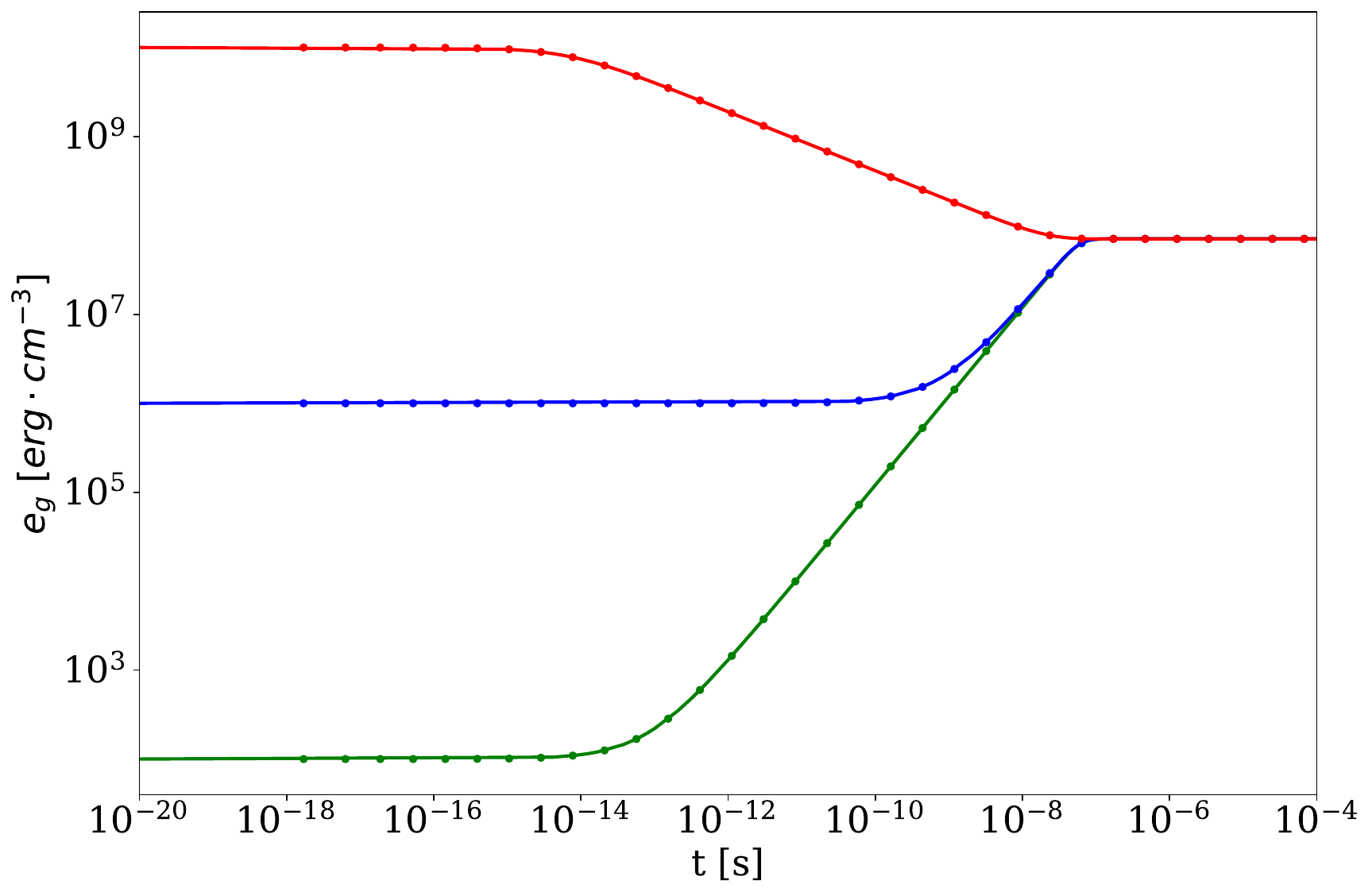}
	\caption{The temporal change of $\eg$ in the radiation and perfect gas coupling tests with three different initial gas conditions.}
	\label{fig:coupling}
\end{figure}

For this test, we choose the same parameters as the ones in \cite{kolb2013}, i.e., $\rho=10^{-7}$ \gcmc, $v=0$, $\kp=0.4$ \cmsg, and $e_{\text{g}}=(3\rho k_{b}\tg)/(2\mu m_{H})$, with $\mu=0.6$ and $m_{H}$ being the mean atomic weight and the mass of a hydrogen atom. To examine the accuracy of the heating and cooling processes of the gas, three initial gas conditions are chosen, they are $e_{\text{g}}=[10^{2},10^{6},10^{10}]$ \ergcmc. Equation \ref{eqn:radmattercoupling} can be integrated numerically by the ode45 function in {\tt Matlab}. We set the initial timestep $dt=10^{-20}$s and increase the timestep by 1$\%$ after each step. The relative error is set $\er=10^{-6}$. The solid lines in Figure \ref{fig:coupling} are the solution from the ode45 and the dots are the numerical solutions of {\tt Guangqi}. The results clearly show excellent agreement between the two groups of solutions.

\subsubsection{Radiation and hydrogen gas coupling}\label{sec:radcouplinghydrogen}

Next, we test the coupling between radiation and matter that has a complex EoS. The design of the test is very similar to the test in Section \ref{sec:perfectcoupling}. We solve Equation \ref{eqn:radmattercoupling} and assume that the gas consists of \ce{H2}, \ce{H}, \ce{H+}, and \ce{e-} and all the species are in LTE. A description of the EoS can be found in \citet{chen2019}. Note that $\eg(\rho,T)$ is now a nonlinear equation of $\rho$ and $T$ (Figure \ref{fig:cv}). We linearize the equation and obtain,
\begin{equation}\label{eqn:radhydrogen}
	C_{V}\pdv{\tg}{t}=-\kp\rho c(a_{R}\tg^{4}-\E)
\end{equation}
where $C_{V}=d\eg/d\tg$ is the heat capacity of the gas. Here, we use a background radiation at constant temperature to heat/cool pure hydrogen gas that will ionize/recombine. The physical conditions of the heating/cooling test are listed in Table \ref{tab:radcoupling}. We set the total time of integration to be $\Delta t=1000$s.
\begin{table}
	\centering
	\begin{tabular}{cccccc}\hline
	type	&	$\trad$     &   $\tg$		&	$\kp$	&	$\rho$	&	$\er$		\\	
			&	(K)	&   (K)   &	(\cmsg)	&	(\gcmc)	&			\\\hline
	heating	&	$10^4$	&  $4000$   &	0.4		&	$10^{-12}$	&	$10^{-10}$		\\
	cooling	&	$4000$	&  $10^4$   &	0.4		&	$10^{-12}$	&	$10^{-10}$	\\\hline
	\end{tabular}
	\caption{From the left to the right: test type; fixed radiation temperature; initial gas temperature; Planck mean opacity of the gas; density of the gas; relative error in the iterative solver.}
	\label{tab:radcoupling}
\end{table}

\begin{figure}
	\centering
	\includegraphics[width=\columnwidth]{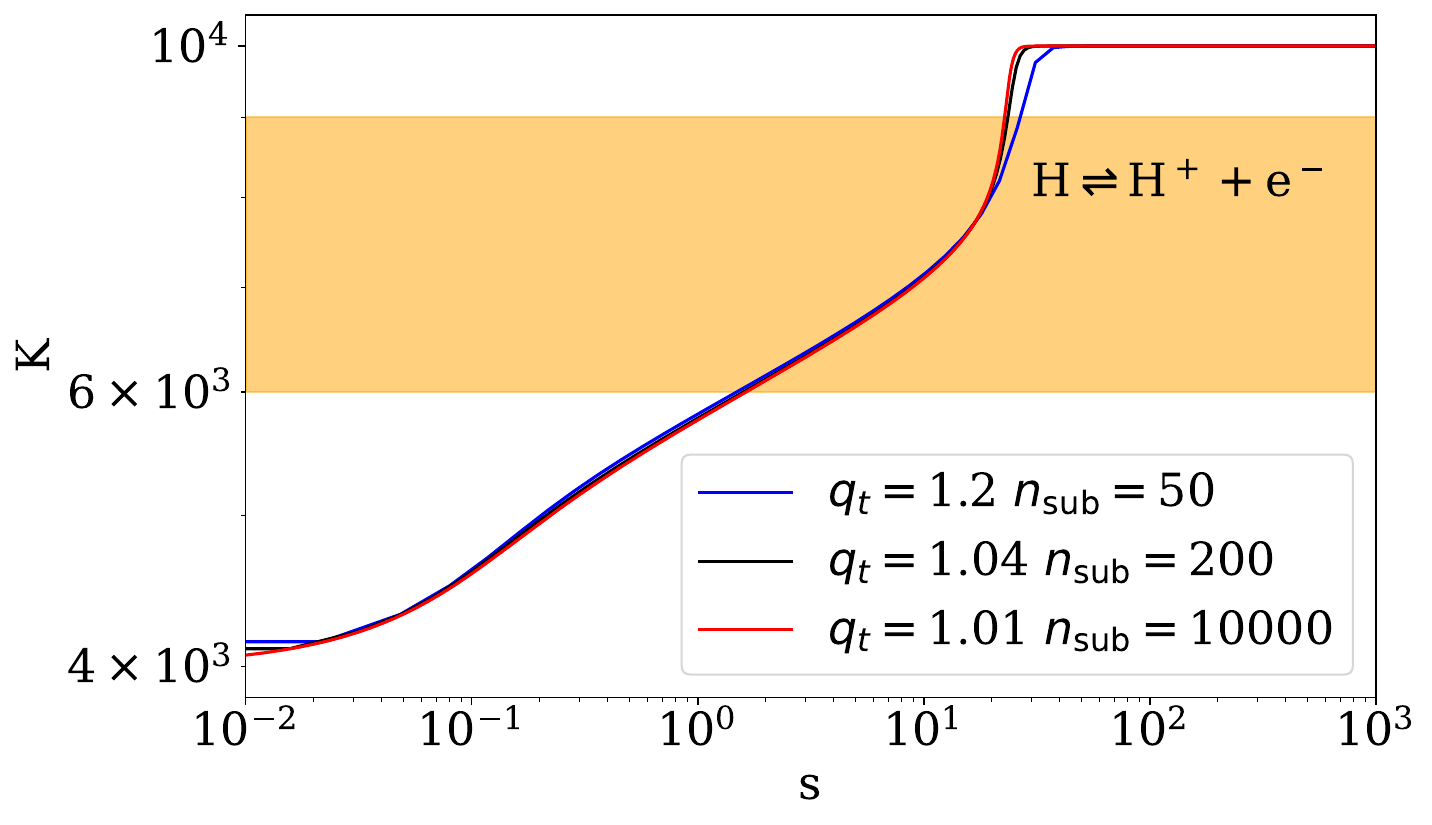}
	\includegraphics[width=\columnwidth]{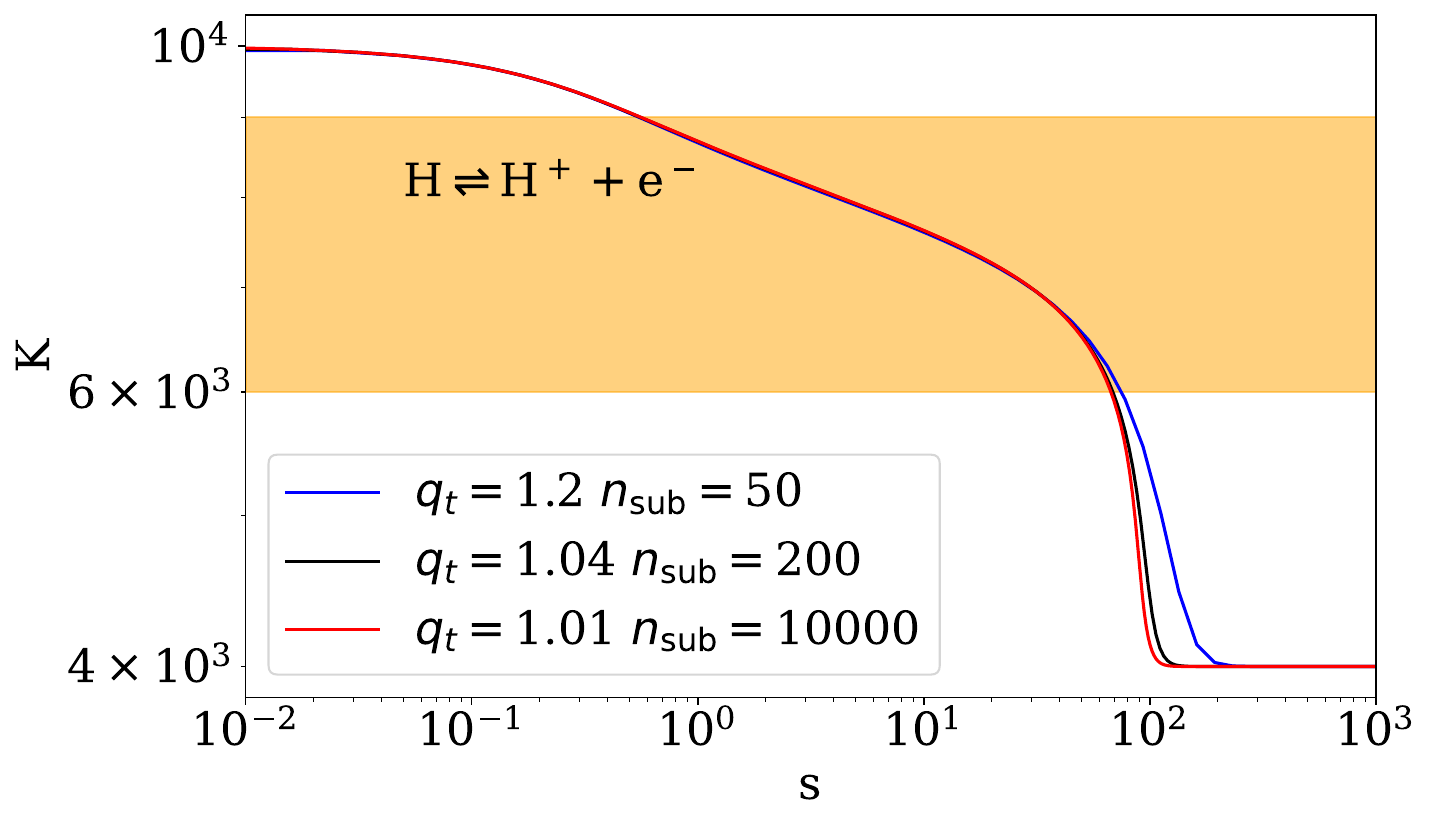}
	\caption{Temporal evolution of $\tg$. Top panel: hydrogen gas heating test; bottom panel: hydrogen gas cooling test. The radiation temperature is held constant at $10^4$K and $4000$K for the heating/cooling test. The initial gas temperature is $4000$K and $10^4$K for the heating/cooling test.}
	\label{fig:radhydrogencoupling}
\end{figure}

In Section \ref{sec:uniformgrid}, we proposed to integrate Equation \ref{eqn:radhydrogen} with an increasing time-step. Figure \ref{fig:radhydrogencoupling} shows the evolution of the $\tg$ with different $q_{t}$ and $n_{\text{sub}}$. The upper and bottom panels show the results of the heating and cooling tests, respectively. We can see that a combination of $q_{t}=1.2$ and $n_{\text{sub}}=50$ combination already gives reasonable results. In practice, the choice of $q_{t}$ and $n_{\text{sub}}$ also depend on the $\kp$ and $\Delta t$ and can be problem dependent.

\subsection{1D tests}

\subsubsection{AMR with a second order general EoS HLLC Riemann solver} \label{sec:hydrotest}

Here we examine the accuracy of the MUSCL scheme general EoS HLLC Riemann solver when combined with the AMR technique for the first time. We use the exact and the approximate Riemann solvers described in \cite{chen2019}.

\begin{figure}[ht!]
	\centering
	\includegraphics[width=\columnwidth]{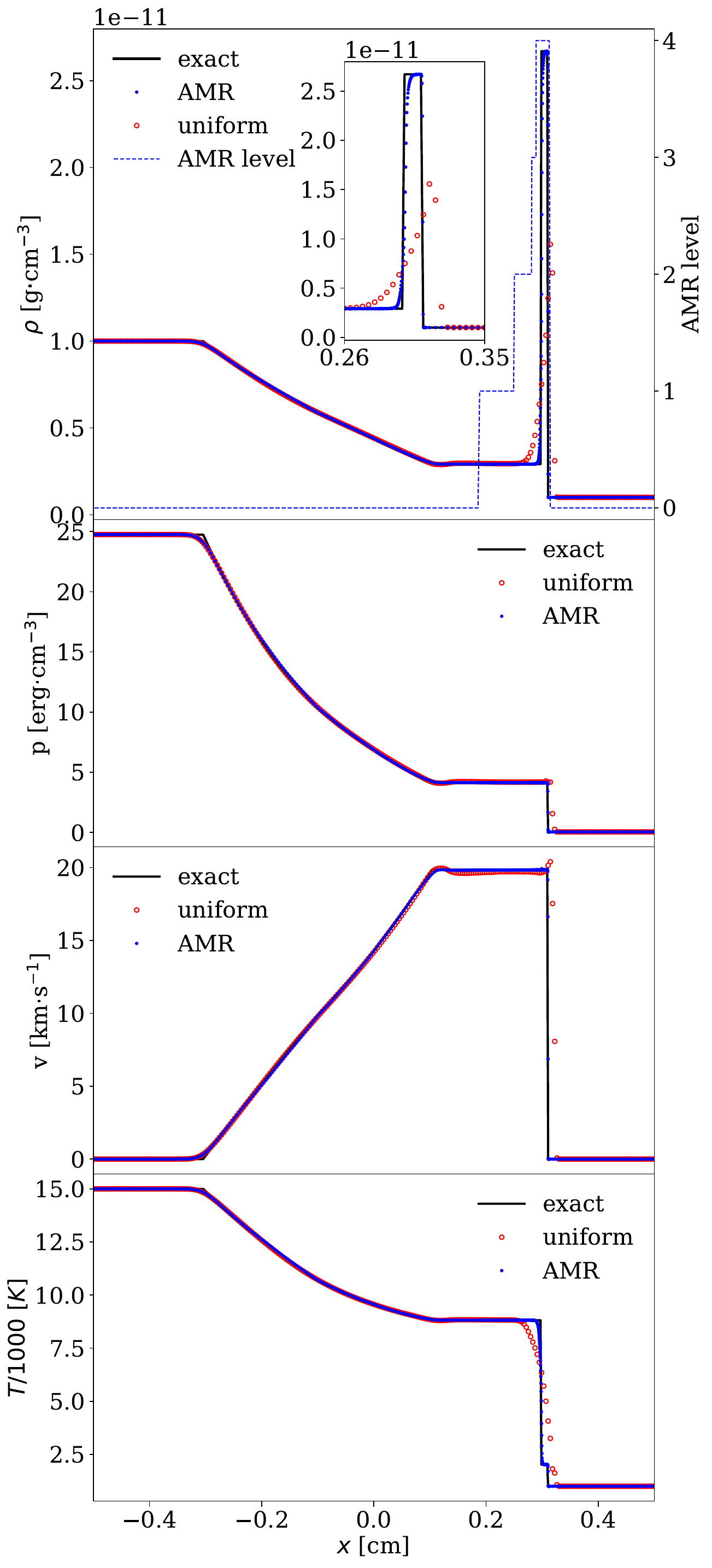}
	\caption{A Sod shock-tube test - with stationary high density and temperature gas on the left and stationary low density and temperature gas on the right - to examine the AMR function with the pure hydrogen EoS. From the top to the bottom, they are the density, pressure, velocity, and temperature profiles. The exact solution is in black solid lines. The blue/red dots show the approximate solution with/without AMR. The AMR level is shown on the secondary $y$ axis with a blue dashed line in the top panel.}
	\label{fig:hydro}
\end{figure}

We setup a Sod shock-tube test to examine the hydrodynamic solver. The Sod shock-tube test has a left initial state $[\rho,v,p,T]=[10^{-11}$ \gcmc, 0 \cms, $2.4753\times10^{1}$ dyn/cm$^{2}$, 15000 K], where the gas is fully ionized. The right initial state is given by $[\rho,v,p,T]=[10^{-12}$ \gcmc, 0 \cms, $4.1256\times10^{-2}$ dyn/cm$^{2}$, 1000 K$]$, where the gas is fully molecular. The CFL number is 0.9 and the simulation time is $1.5\times10^{-7}$ s. The computational domain is $[-0.5,0.5]$ cm and $x=0$ separates the left and right states. Figure \ref{fig:hydro} shows the density, velocity, pressure, and temperature profiles calculated by {\tt Guangqi} and the exact solution from \cite{chen2019}. The base resolution has $n_{x}=256$ grid cells.
The red hollow circles show results obtained from the base resolution, while the blue dots correspond to results with 4 levels of AMR.
The refinement criterion of this test is
\begin{equation}
	\frac{2|\rho_{i+1}-\rho_{i-1}|}{\rho_{i}}\ge0.02,
\end{equation}
which is mostly concentrated around the shock.

We first notice that beyond the shock region, the results with and without AMR both agree very well with the exact solution, including in the rarefaction wave, testifying the accuracy of our numerical scheme in general. The shock front and contact discontinuity are not properly resolved in uniform grid, but are well captured when AMR is enabled. Therefore, the AMR technique improves both accuracy and efficiency.

\subsubsection{Dynamical radiative diffusion}

In an extremely opaque environment, e.g., deep inside a star, radiation diffusion may be less efficient in transporting energy than advection. This situation, known as radiation trapping, occurs when the diffusion timescale is much longer than the advection timescale. Here, we test the coupling between the advection and diffusion parts of the radiation hydrodynamic solver of {\tt Guangqi}, by solving a 1D problem with background fluid motion and radiative diffusion while turning off radiation-matter coupling \citep{jiang2021}. In the absence of the background fluid motion, the test will reduce to the normal diffusion problem \citep{commercon2011,kolb2013}.
The computational domain is $[-2,14]$ cm with $N=1024$. We set uniform fluid conditions with $v=1.8$ \kms, $\tg=0.1$ K, and its opacities are $\rho\kr=10^{7}$ cm$^{-1}$, and $\kp=0$.
The density of the fluid does not matter here because the fluid and the radiation are not coupled ($\kp=0$). We run the simulation for $t=6.2\times10^{-5}$ s and set the CFL number to be 1 and $\er=10^{-8}$. 

We set the initial condition for radiation energy density to be,
\begin{equation}
	\E=\begin{cases}
		E_{0}/(2\Delta x),&\quad|\text{center}(x)|<\Delta x		\\
		1,&\quad\rm{elsewhere}
		\end{cases}
\end{equation}
where $\text{center}(\cdot)$ is a function that returns the cell center's coordinate, i.e., the central two cells neighboring $x=0$ satisfy the first condition. The total energy of the central energy packet is set to $E_0=2\times10^5$ erg.
Both the left and right radiation boundary conditions are zero-gradient, i.e., $\partial\E/\partial r=0$. The fluid boundary condition is fixed at $v=1.8$ \kms and $\tg=0.1$ K.

It is straightforward to obtain the analytic solution of $\E$ as
\begin{equation}
	E_{a}(x,t)=1+\frac{E_{0}}{\sqrt{4D\pi t}}\exp{\frac{-(x-vt)^{2}}{4Dt}},
\end{equation}
where $D=c/(3\rho\kr)$. Over the duration of the test, we can calculate the characteristic diffusion length $l_{\rm diff}=\sqrt{4Dt}\approx0.498$ cm, which is much shorter than the hydrodynamical length $l_{\rm dyn}=vt=11.16$ cm.

\begin{figure}[ht!]
	\centering
	\includegraphics[width=\columnwidth]{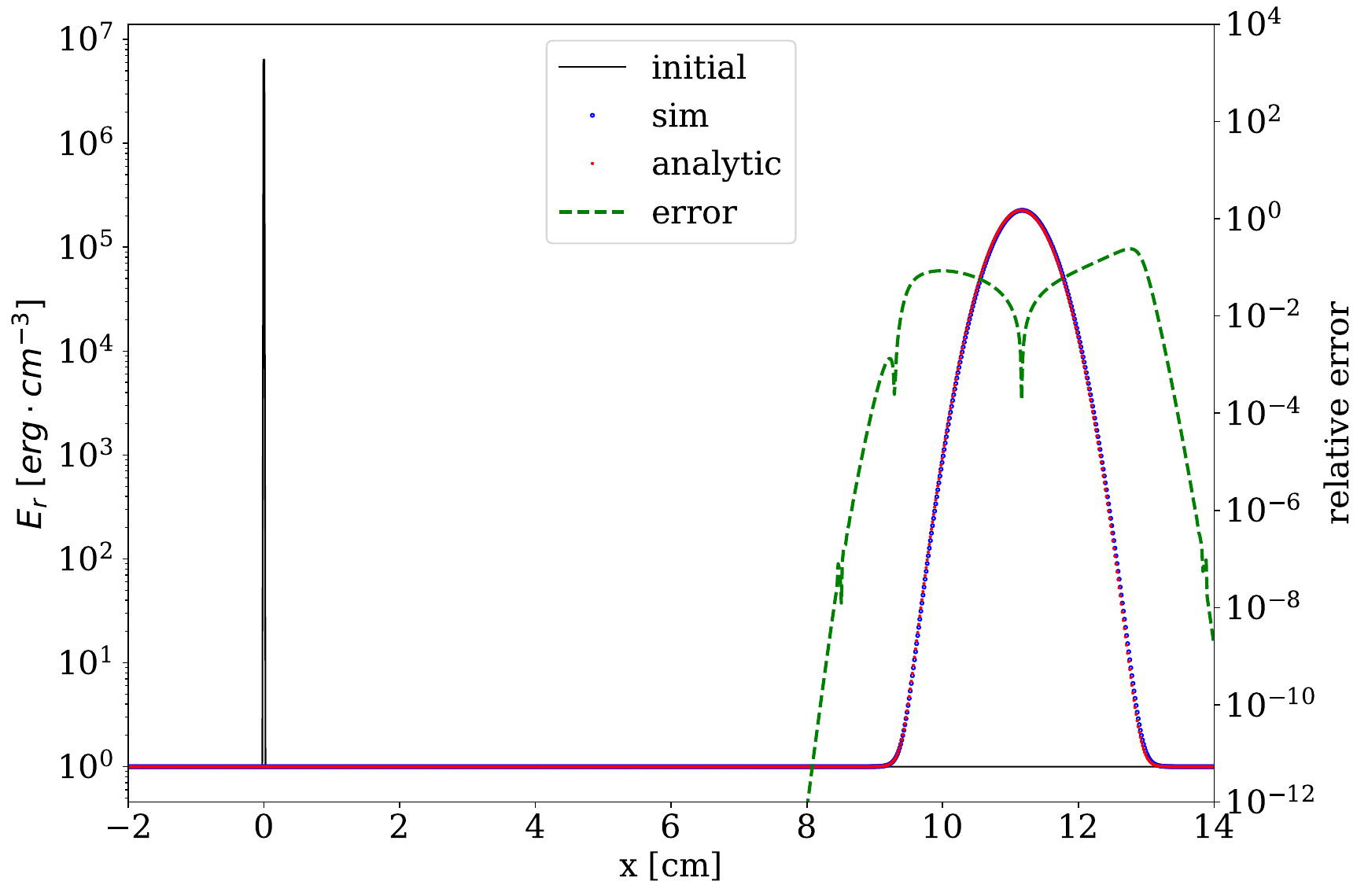}
	\caption{Dynamical diffusion test result at $t=6.2\times10^{-5}$s. Black line: the initial condition. Red and blue dots: the analytic solution and the simulated result respectively. Green dashed line: the relative error.}
	\label{fig:diffusion}
\end{figure}

Figure \ref{fig:diffusion} shows the initial condition, analytic solution, numerical solution, and relative error, defined as
\begin{equation}
    \text{relative error}=\frac{|E_{a}-\E|}{E_{a}},
\end{equation}
where $\E$ is the numerical solution.
As we can see, the energy packet diffuses and advects well follows the analytical solution, and the error is kept to be relatively small. Note that in this test,
advection is much more significant than diffusion
given $l_{\rm diff}<l_{\rm dyn}$.

\subsubsection{Radative shocks with a perfect gas EoS}\label{sec:perfectradshock}

In this subsection, we test the full radiation hydrodynamic solver by studying radiative shocks. We follow the numerical setup of \cite{ensman1994} to construct the sub-critical and supercritical radiative shocks. The same setup has been adopted by \cite{commercon2011,kolb2013,colombo2019a}.

The physical picture of these simulations is that the gas runs into a hard and non-conducting wall and radiates away its energy in the shock. Our computational domain is $[0,7\times10^{10}]$ cm with a base resolution of $N=256$. One difference in this work is that we apply 3 levels of SMR that can cover the shock throughout the whole simulation. Therefore, the effective resolution is still $2048$, which was adopted by all the aforementioned works. The gas has a uniform density of $\rho=7.78\times10^{-10}$ \gcmc\ and a temperature $\tg=10$K. The initial gas velocity is $v_{\text{g}}=-6$ \kms\ and $v_{\text{g}}=-20$ \kms\ for the sub-critical and super-critical radiative shock tests, 
and we sample the results at $t=38000$ s and $t=7500$ s, respectively. The gas is a $\gamma$-law gas with $\gamma=1.4$ and $\mu=1$. We adopt constant photon mean free paths with $\rho\kr=\rho\kp=3.1\times10^{-10}$ cm$^{-1}$
(implying that the opacity $\kp$ and $\kr$ are set to be inversely proportional to density). 
The initial radiation energy density is $\E=a_{R}\tg^{4}$. The CFL number is 0.4 and the relative error tolerance of the matrix solver is set to $\er=10^{-5}$. The left boundary is reflective for the fluid variables and zero-gradient for radiation. The right fluid boundary is fixed and the right radiation boundary is also zero-gradient. The fixed right fluid boundary has the same state as the corresponding initial state of the fluid.

\begin{figure}[ht!]
	\centering
	\includegraphics[width=\columnwidth]{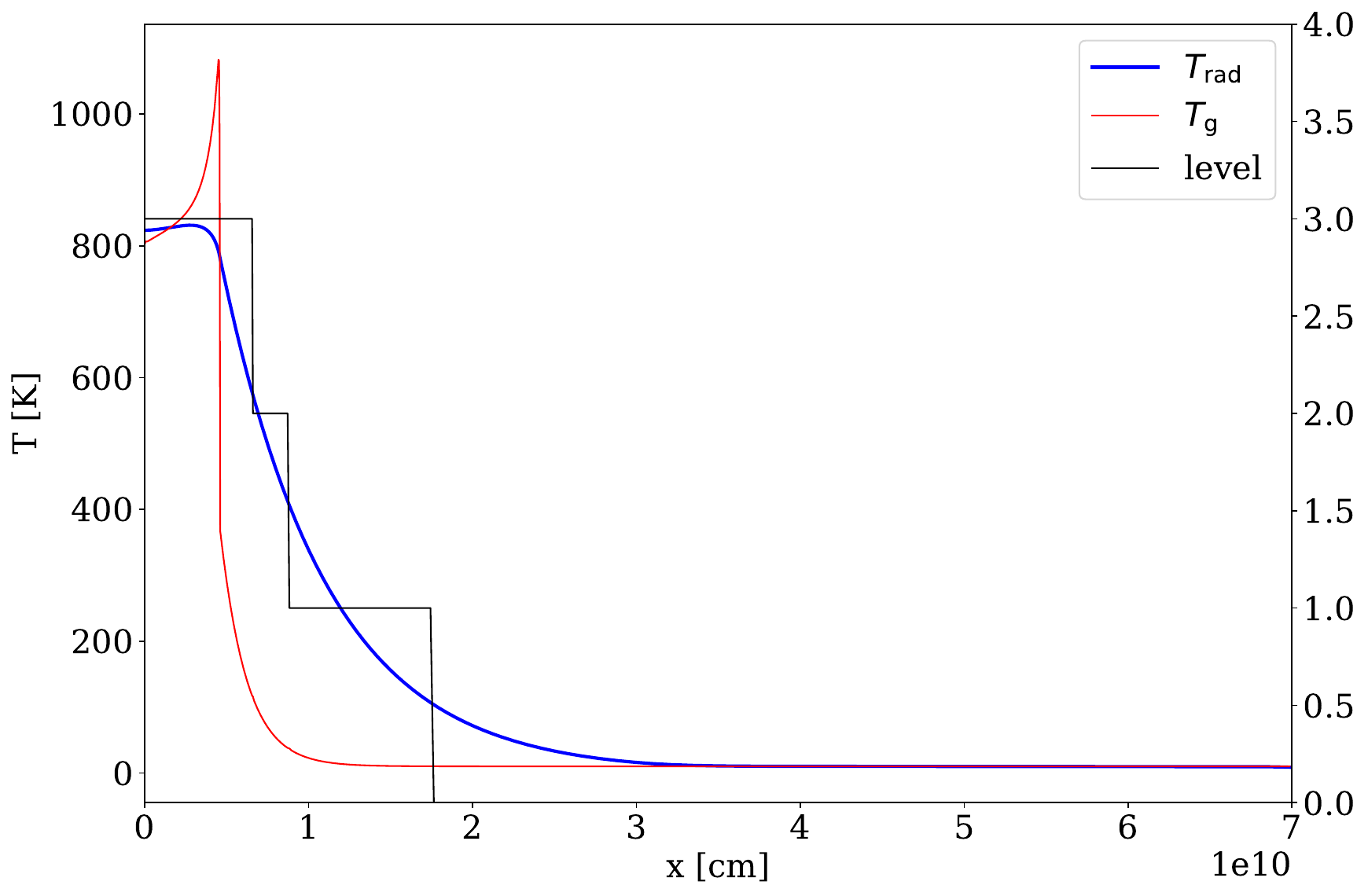}	\\
	\includegraphics[width=\columnwidth]{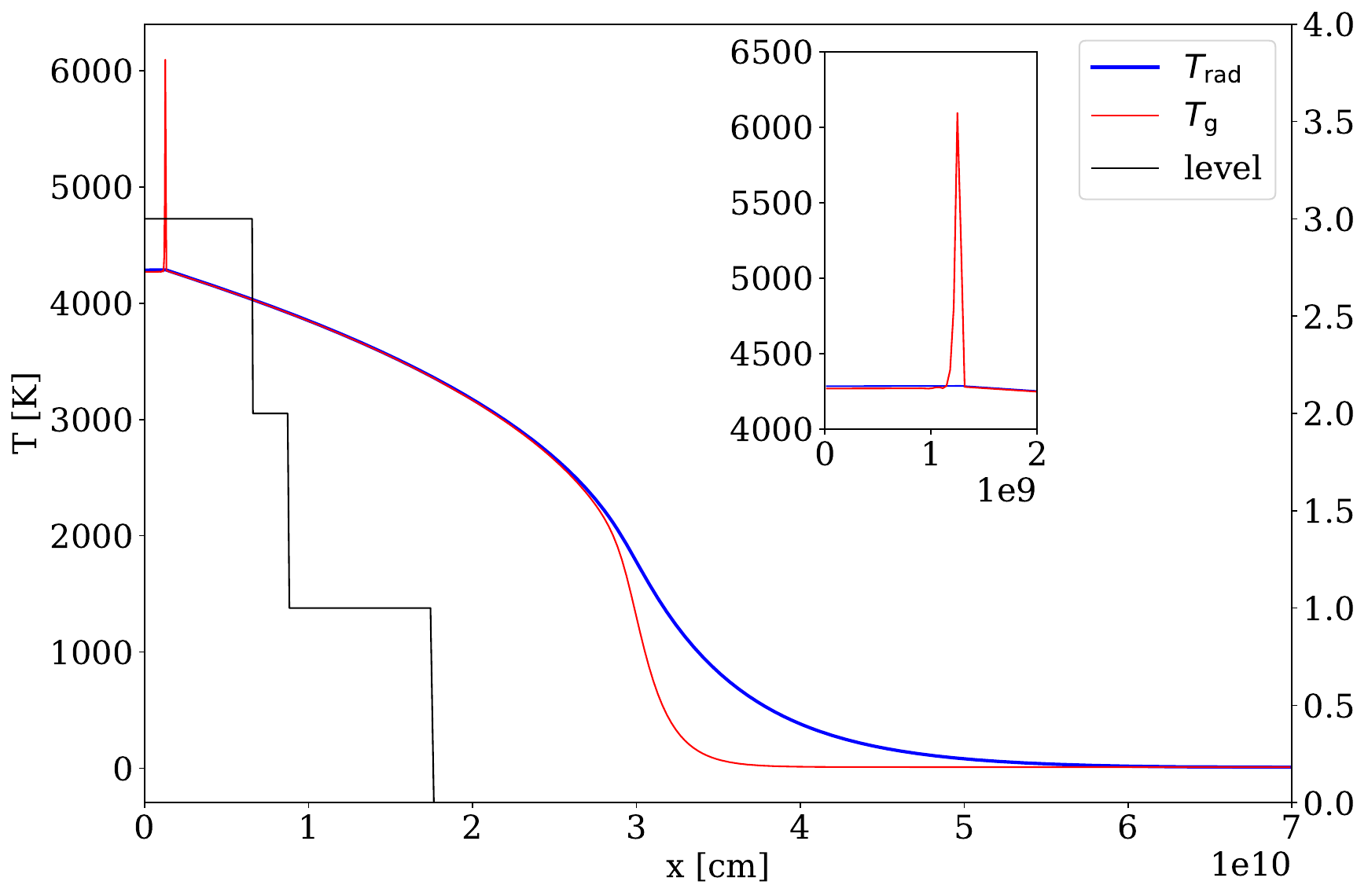}
	\caption{Top panel: sub-critical shock test with 3 levels of SMR applied to $[0,4\times10^{9}]$ cm. Bottom panel: supercritical shock test with 3 levels of SMR applied to $[0,4\times10^{9}]$ cm, a zoom-in plot focuses on the Zel'dovich spike. Red line: the gas temperature. Blue dots: the radiation temperature. Black solid line: the SMR level.}
	\label{fig:rhdshock}
\end{figure}

\begin{table*}
	\centering
	\begin{tabular}{ccccc}\\ \hline
		Temp		&	Analytic		&	This work	&	\cite{kolb2013}	&	\cite{colombo2019a}	\\	\hline
		$T_{2}$	&	$\approx$865 K	&	816.7 K	&	816.6 K	&	817 K	\\
		$T_{-}$	&	$\approx$315 K 	&	321.7 K	&	331.9 K	&	332 K	\\
		$T_{+}$	&	$\approx$1075 K	&	1068.1 K 	&	1147.1 K 	&	1151 K	\\	\hline
	\end{tabular}
	\caption{Comparison of the temperature at different regions of the shock between analytical estimates and numerical results
    %analytic estimate temperatures to simulations 
    in the sub-critical radiative shock test (see Section \ref{sec:perfectradshock}). The second column is the analytic estimates. Numerical results are obtained from this work and other literature is shown in the next three columns.}
	\label{tab:rhdcompare}
\end{table*}

Figure \ref{fig:rhdshock} shows the profiles of $\tg$ and $\trad$ of the sub-critical and super-critical shocks. In the case of the sub-critical radiative shock, $\trad>\tg$ in the pre-shock region and $\tg>\trad$ in the post-shock region, indicating that the shock strength is not strong enough to preheat the upstream, and the shocked gas carries excess energy to the downstream. Three characteristic temperatures can be estimated analytically, which are the final equilibrium post-shock temperature \citep{mihalas1984,ensman1994},
\begin{equation}
    T_{2}\approx\frac{2(\gamma-1)v_{\text{g}}^{2}}{R_{G}(\gamma+1)^{2}},
\end{equation}
the temperature immediately ahead of the shock front \citep{mihalas1984},
\begin{equation}
    T_{-}\approx\frac{\gamma-1}{\rho vR_{G}}\frac{2\sigma_{\rm sb}T^{4}_{2}}{\sqrt{3}},
\end{equation}
and the temperature immediately behind the shock front \citep{mihalas1984},
\begin{equation}
    T_{+}\approx T_{2}+\frac{3-\gamma}{\gamma+1}T_{-},
\end{equation}
where $R_{G}=k_{b}/\mu m_{H}$. Table \ref{tab:rhdcompare} lists the analytically estimated temperature and the numerical results of three works. Our results are in agreement with the analytic estimations.

In the case of super-critical shock, we have $\trad=\tg$ in the pre-shock and post-shock region, indicating that the shock is strong enough to preheat the upstream and cools down to the equilibrium temperature in a very short distance in the downstream. The very short distance of cooling is the Zel'dovich spike.

\subsubsection{Radiative shocks with the hydrogen EoS}\label{sec:rhdeos}

This test represents the first comprehensive numerical benchmark that integrates radiative transfer, complex EoS physics, and hydrodynamics. The setup is fundamentally similar to the supercritical shock test described in Section \ref{sec:perfectradshock}; however, in this configuration, the shock front preheats the upstream material sufficiently to trigger hydrogen ionization and dissociation. We highlight the specific modifications to the benchmark setup below:
\begin{enumerate}
\item To ensure the shock is capable of preheating the upstream gas to an ionized state, the gas velocity is set to $v_{\text{g}} = -70$ km s$^{-1}$. The simulation is evolved until $t = 2500$ s.
\item The upstream progenitor gas is initialized in a purely molecular state.
\item For the radiation field at the right boundary, we adopt the Milne boundary condition \citep[see Chapter 11 of][]{castor2004}:
\begin{equation}\label{eqn:milne}
\E=-\frac{1}{3/2\rho\kr}\pdv{\E}{x},
\end{equation}
which physically represents a blackbody absorber at the outer boundary.
\item We set $\epsilon_{r} = 10^{-8}$.
\end{enumerate}

We employ the hydrogen EoS and the analytical EoS solver as described in \citet{chen2019}, which assumes chemical equilibrium among $\text{H}_2$, $\text{H}$, $\text{H}^+$, and $e^-$. To ensure a converged solution for the radiation-thermodynamic subsystem, we implement sub-cycling with a stretching factor of $q_{t} = 1.4$. We tested various sub-cycle counts, $n_{\text{sub}} = \{1, 5, 6\}$, and found that the results for $n_{\text{sub}} = 5$ and $n_{\text{sub}} = 6$ are nearly identical, indicating convergence.

\begin{figure}
    \centering
    \includegraphics[width=\columnwidth]{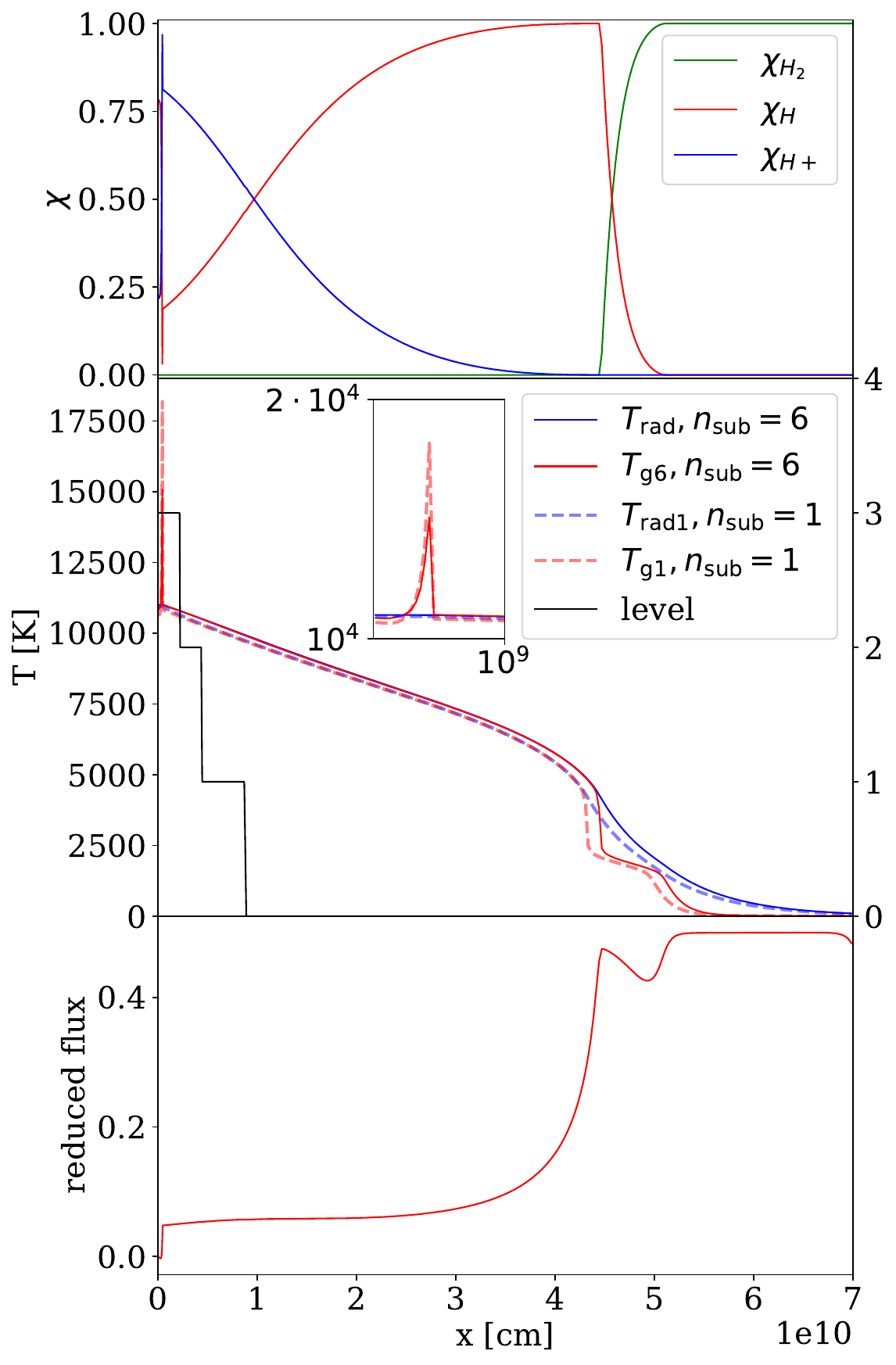}
    \caption{Top panel: the mass fraction of hydrogen species. Middle panel: $\tg$ and $\trad$ of solutions with $q_{t}=1.4$ and $n_{\text{sub}}=\{1,6\}$. The dashed lines are the unresolved solutions ($n_{\text{sub}}=1$) and the solid lines are the converged solutions ($n_{\text{sub}}=6$). We apply 3 levels of SMR to $[0,2\times10^{9}]$cm region. Bottom panel: reduced flux $|\F|/(cE_{r})$.}
	\label{fig:radhydrogenshock}
\end{figure}

Figure \ref{fig:radhydrogenshock} shows the results of the radiative shock with a hydrogen EoS. The solid lines are the converged solutions, and for comparison, we show the solution of $n_{\text{sub}}=1$ in dashed lines. The top panel shows the mass fraction of different hydrogen species, defined as,
\begin{equation}
    \chi_{s}=\frac{n_{s}m_{s}}{\rho},
\end{equation}
where $s$ represents any species. The overall temperature profile is similar to the super-critical shock simulation of the $\gamma$-law gas, in which $\tg=\trad$ in the upstream and downstream of the shock. Due to the coupling between the radiation hydrodynamics and EoS, hydrogen is ionized at the Zel'dovich spike. From the shock to the upstream region, the ionization fraction gradually decreases as $\tg$ drops. Consequently, $\chi_{\ce{H}}$ increases. In the far upstream, $\chi_{\ce{H}}$ drops rapidly between $[4\times10^{9},5\times10^{9}]$cm because \ce{H2} is not dissociated as $\tg$ drops. In the middle panel, $\tg$ and $\trad$ profiles are similar to Fig \ref{fig:rhdshock} except that there is an additional kink in $\tg$ due to the hydrogen dissociation. In comparison, the unresolved solution ($n_{\text{sub}}=1$) overestimates the shock temperature. The bottom panel shows the reduced flux $\F/(c\E)$, note that the flux is not zero at the right boundary as we allow the radiation to leave the computational domain.

A closely related application of this test is the study of accreting gas giants \citep{chen2022} (in a spherical coordinate), where low-angular momentum gas may fall directly onto the atmosphere of the gas giants. Luminous shocks may form at the top layer of their atmosphere and dissociate \ce{H2} of the incoming gas.

\subsection{2D tests}\label{sec:2dtests}

\subsubsection{Kelvin-Helmholtz instability}\label{sec:khi}

The Kelvin-Helmholtz instability (KHI) test is a standard hydrodynamic test in 2D. We follow \cite{stone2020,wibking2022} to set up the test, where we turn off radiation transport and use Van Leer slope limiter. The boundary is periodic in both $x$ and $y$ directions. The computational domain is $[-0.5,0.5]\times[-0.5,0.5]$. The initial condition of the simulation is
\begin{eqnarray}
    \rho&=&1.5-0.5\tanh\bigg(\frac{|y|-0.25}{L_1}\bigg),   \\
    v_{x}&=&0.5\tanh\bigg(\frac{|y|-0.25}{L_1}\bigg),  \\
    v_{y}&=&A\cos(4\pi x)\exp\bigg[-\frac{(|y|-0.25)}{L_2^{2}}\bigg].
\end{eqnarray}
with the shearing layer thickness $L_1=0.01$, $L_2=0.2$, and perturbation amplitude $A=0.01$. The initial pressure is uniform with $p=2.5$ and we adopt an adiabatic index $\gamma=1.4$. We run the simulation with both uniform high resolution and AMR. The uniform high resolution is $2048\times2048$, and the base level of the AMR simulation has a resolution of $256\times256$ with 3 levels of mesh refinement, thus the effective highest resolution is also $2048\times2048$. The refinement criteria are
\begin{eqnarray}
    (\Delta x\|\nabla v_{x}\|)/v_{x}&>&0.01,\quad\text{or}\\
    (\Delta x\|\nabla v_{y}\|)/v_{y}&>&0.01,
\end{eqnarray}
and the derefinement criteria are,
\begin{eqnarray}
    (\Delta x\|\nabla v_{x}\|)/v_{x}&<&0.005,\quad\text{and}\\
    (\Delta x\|\nabla v_{y}\|)/v_{y}&<&0.005,
\end{eqnarray}
We choose the CFL number to be 0.4 and stop the simulation at $t=1.2$. The simulations are run on an AMD EPYC 7763 CPU with 32 cores.

\begin{figure*}
    \centering
    \includegraphics[width=0.325\textwidth]{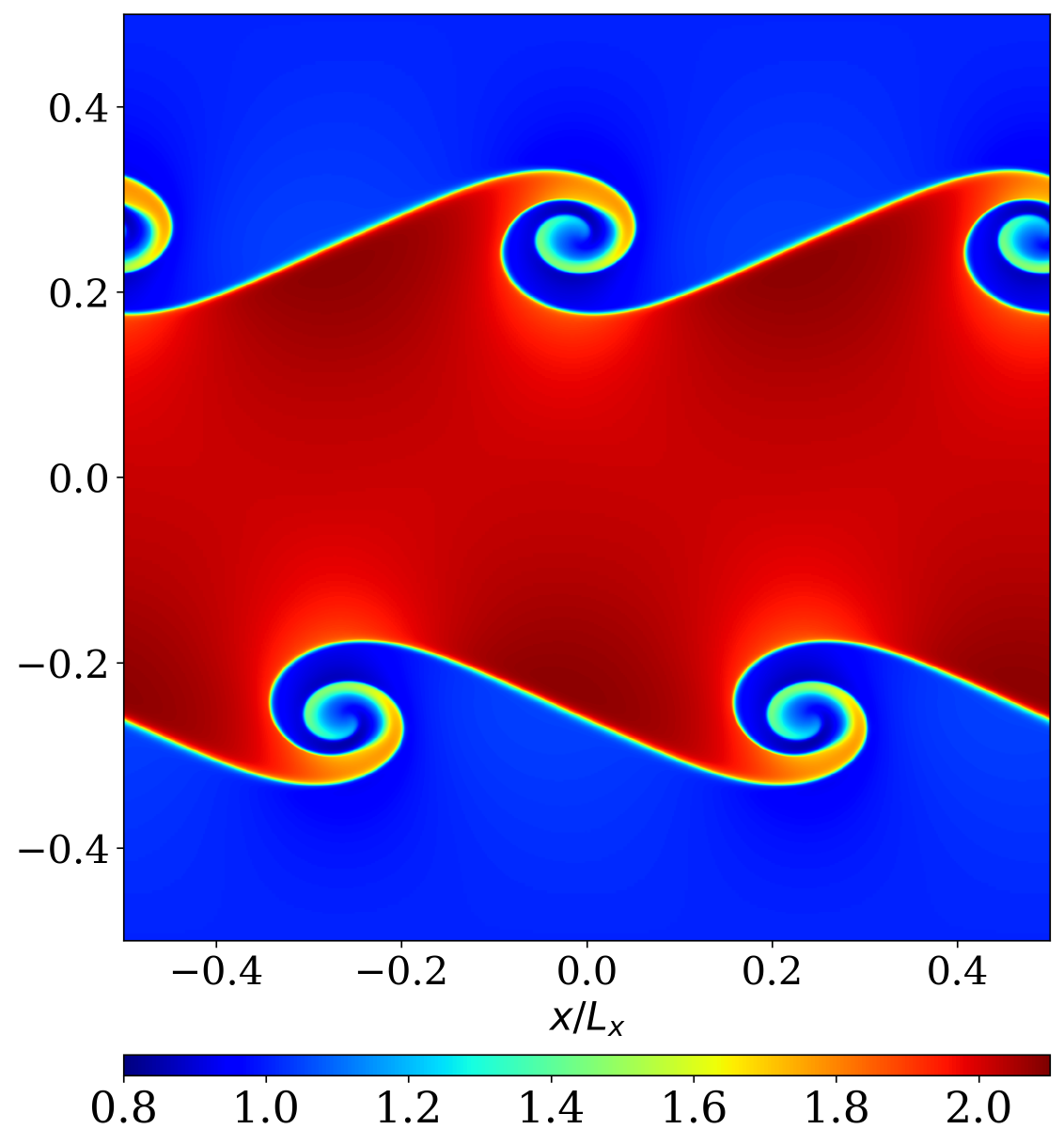}
    \includegraphics[width=0.325\textwidth]{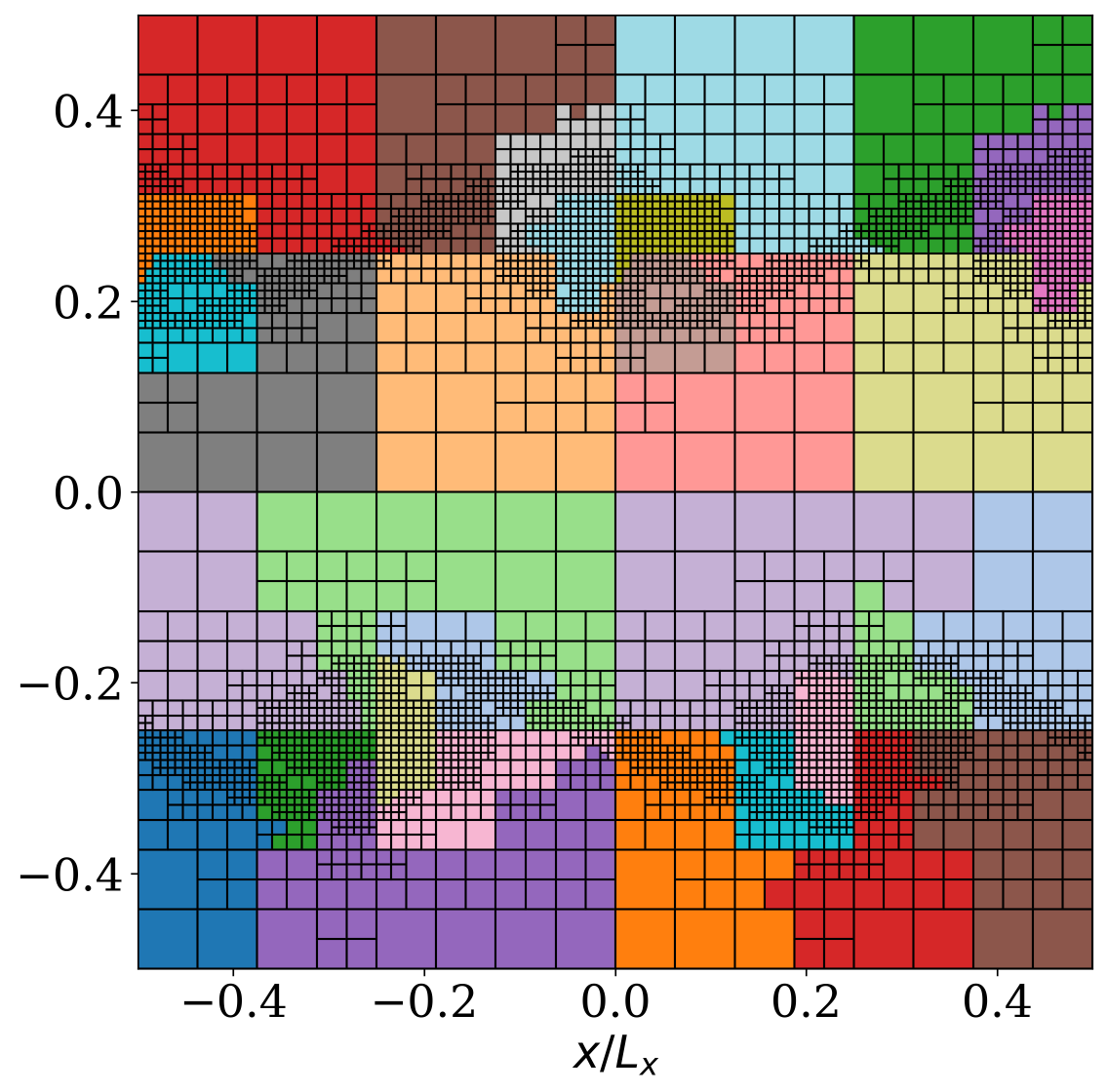}
    \includegraphics[width=0.325\textwidth]{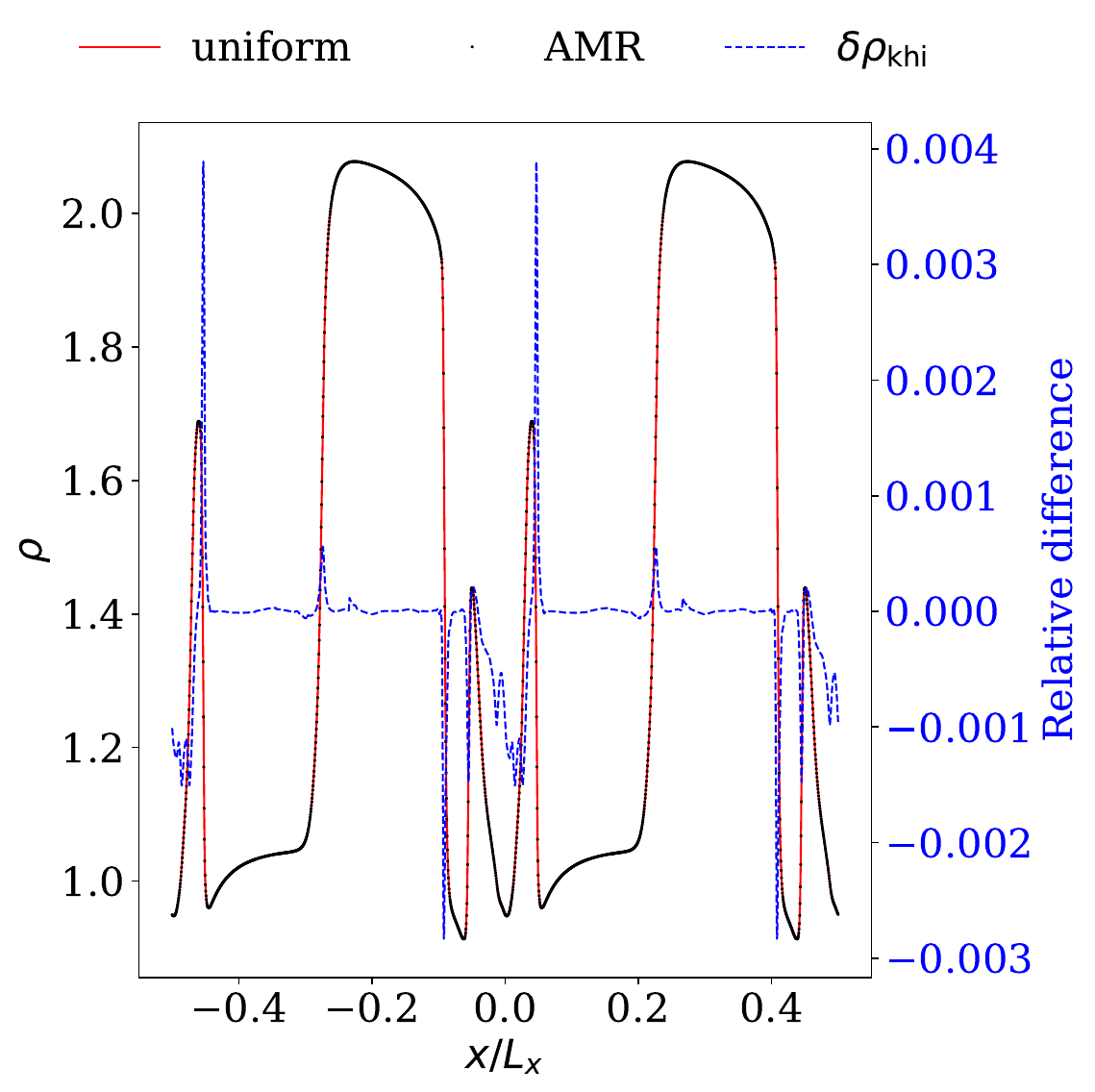}
    \caption{Results of the Kelvin-Helmholz instability test at $t=1.2$. Left panel: the density of the AMR simulation. Middle panel: the AMR block structure and domain decomposition into 32 cores colored according ``tab20" colormap periodically. Right panel: the density profile at $y=0.25$ and $t=1.2$, the red line and black dots show the simulation results of fixed resolution ($2048\times2048$) and AMR ($256\times256$ with 3 levels of mesh refinement).}
    \label{fig:khi}
\end{figure*}

We present the results in Figure \ref{fig:khi}. The density structure on the left panel is similar to the results from other works. The middle panel shows that the vortices are resolved by high resolution while the interior of the streams does not require high resolution. The domain decomposition into 16 cores is according to the Hilbert curve, and each sub-domain is connected. We compare the AMR simulation to uniform high resolution simulation by plotting their density profile at $t=1.2$ and $y=0.25$, where the density gradient is the steepest. On this horizontal line, the AMR simulation has low resolution and high resolution regions but the difference between the red line (fixed high resolution) and black dots is indistinguishable. Quantitatively, we interpolate the AMR density on the uniform grid to calculate their relative difference by,
\begin{equation}
    \delta\rho_{\rm{khi}}=(\rho_{\rm{AMR}}-\rho_{\rm{uniform}})/\rho_{\rm{uniform}}.
\end{equation}
The relative difference is within $0.4\%$, similar to the results in other works.

\subsubsection{Performance of PSAMA}\label{sec:angularmomentumtest}

To evaluate the robustness and performance of the PSAMA scheme (Section \ref{sec:angular}), we utilize an isolated, inviscid, and locally isothermal hydrostatic thin-disk model. In spherical polar coordinates, the governing Euler equations for this hydrostatic configuration are given by:
\begin{eqnarray}
    \frac{v_{\phi}^2}{r}&=&g+\frac{1}{\rho}\pdv{p}{r},   \label{eqn:diskvr}\\
    v_{\phi}^2\cot\theta&=&\frac{1}{\rho}\pdv{p}{\theta}, \label{eqn:diskvtheta}
\end{eqnarray}
where $g=GM_{s}/r^2$ is the central gravity, $p$ is the ideal gas pressure. We can obtain $v_{\phi}$ from Equation \ref{eqn:diskvr},
\begin{equation}
    v_{\phi}^{2}=rg+T\bigg(\pdv{\ln T}{\ln r}+\pdv{\ln\rho}{\ln r}\bigg).    \label{eqn:diskvphi}
\end{equation}

To make it an isolated disk in a nearly vacuum circumstance, we assume $\rho$ and $T$ to have the following dependence on $r$ and $\theta$,
\begin{eqnarray}
    \rho&=&\rho_0\bigg(\frac{r}{r_0}\bigg)^a \sig_{1}(r)\sig_2(r) f(\theta) \label{eqn:diskrho}\\
    \sig_1(r)&=&(1+\exp(-d_1 (r-r_{\text{in}})/r_{\text{in}}))^{-1}   \\
    \sig_2(r)&=&(1+\exp(d_2 (r-r_{\text{out}})/r_{\text{out}}))^{-1}    \\
    T&=&T_0\bigg(\frac{r}{r_0}\bigg)^b=\frac{p_0}{\rho_0}\bigg(\frac{r}{r_0}\bigg)^b, \label{eqn:disktemp}
\end{eqnarray}
where $\sig_1(r)$ is to create an inner cavity, $\sig_2(r)$ is to truncate the outer disk, $r_{\text{in}}$, $r_{\text{out}}$, $d_1$, and $d_2$ controls the location of the inner cavity, truncation radius, gradient of the density slopes, respectively. In addition, $r_0$, $\rho_0$, $T_0$, $p_0$, $a$, and $b$ are constants that create the power-law part of the disk profiles. Equation \ref{eqn:diskvphi}-\ref{eqn:disktemp} characterizes an isolated hydrostatic disk. 
To calculate $f(\theta)$, we solve Equation \ref{eqn:diskvtheta},
\begin{equation}
    \frac{v_{\phi}^{2}}{T}=\pdv{\ln\rho}{\ln\sin\theta}.
\end{equation}
As a result,
\begin{equation}
    f(\theta)=(\sin\theta)^{v_{\phi}^2/T}.  \label{eqn:ftheta}
\end{equation}

In a hydrostatic disk, the only non-zero physical variables are the azimuthal velocity $v_{\phi}$, density $\rho$, and temperature $T$, as defined by Equations \ref{eqn:diskvphi}, \ref{eqn:diskrho}, and \ref{eqn:disktemp}, respectively. When initialized in a steady state, the analytical profile should relax to a numerical equilibrium that remains close to the initial condition. However, Equation \ref{eqn:ftheta} implies that the polar regions will reach extremely low densities. To handle this numerically, we implement a density floor during initialization. A sufficiently low floor density ensures that the ambient environment approximates a vacuum, thereby minimizing its dynamical impact on the disk.

For this validation test, we adopt the following dimensionless parameters: $M_sG=1$, $r_0=1$, $r_{\text{in}}=4$, $r_{\text{out}}=16$, $d_1=15$, $d_2=30$, $\rho_0=1$, $T_{0}=0.0022$, $p_{0}=0.0022$, $a=-2$, and $b=-1$. Thus, the disk scale height is $H/R=0.047$. We set the simulation domain to be $r\in[1,40]$ and $\theta\in[0,\pi/2]$. The resolution of the simulation is $512\times128$. The mesh is stretched (non-uniform grid) in the $r$ coordinate with a ratio of $q_{r}=1.0058682$ ($\Delta r_{i+1}=q_{r}\Delta r_{i}$, $\Delta r_{i}$ is the $i$th cell size) so that the spacing is logarithmic and uniform in the $\theta$ coordinate.

We set the CFL number to 0.5 and apply mirror symmetry boundary conditions at both the equator and the pole. The inner and outer radial boundaries are maintained at constant density and temperature with zero rotation: $\rho_{\text{in}}=10^{-6}$, $\rho_{\text{out}}=10^{-18}$, $T_{\text{in}}=0.0022$, and $T_{\text{out}}=5.5 \times 10^{-5}$. During initialization, if the density calculated by Equation \ref{eqn:diskrho} falls below $\rho_{\text{out}}$, we set $\rho = \rho_{\text{out}}$ and the angular momentum to zero. Notably, we do not enforce a density floor during the subsequent hydrodynamic evolution, as the combination of the CFL condition and the PSAMA scheme ensures density positivity. The extremely low-density ambient gas ($\rho_{\text{out}}/\rho_{0} = 10^{-18}$) with zero angular momentum eventually infalls from the outer boundary; however, its low inertia ensures a negligible impact on the global evolution of the disk.

\begin{figure*}
    \centering
    \includegraphics[width=0.495\textwidth]{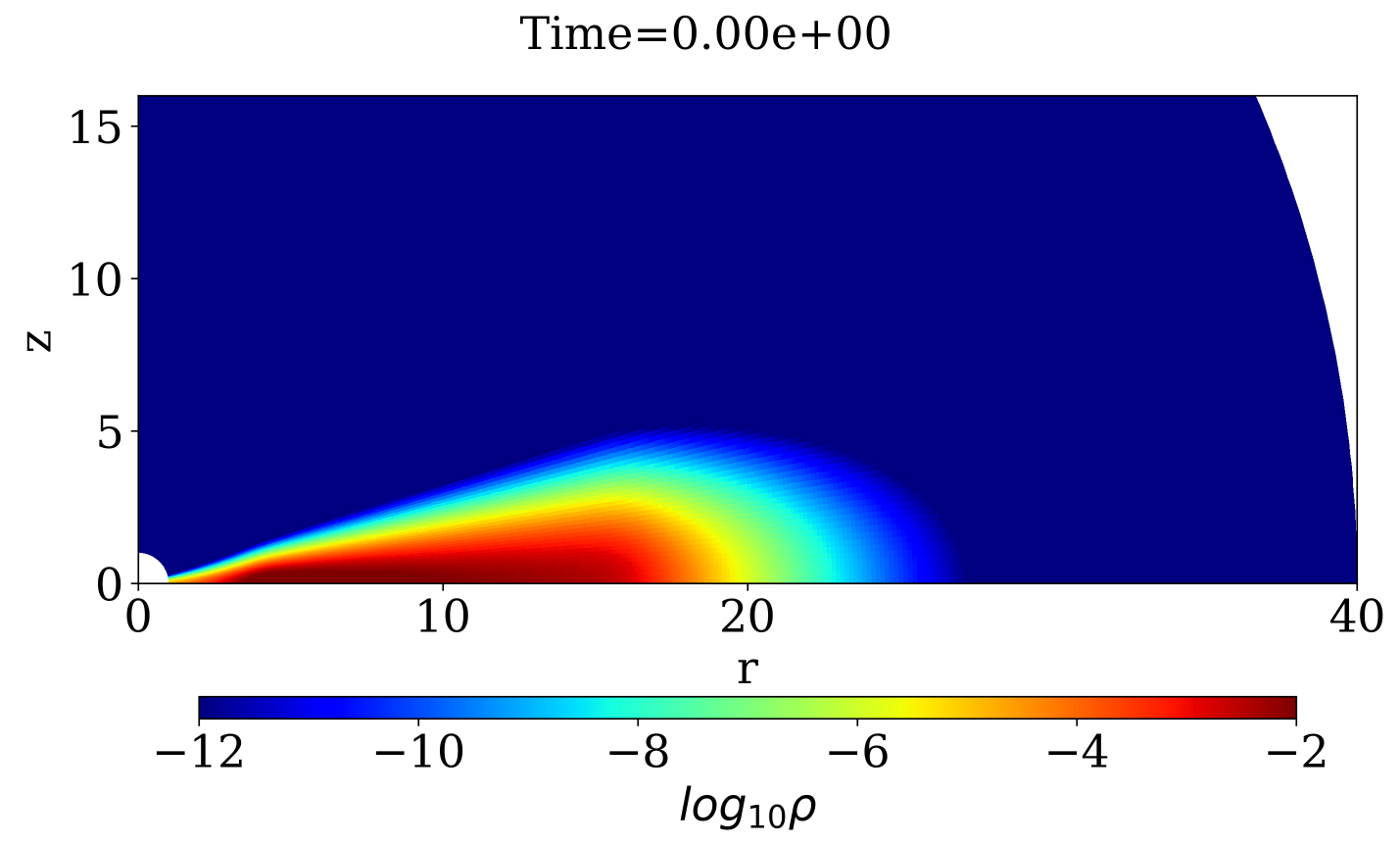}
    \includegraphics[width=0.495\textwidth]{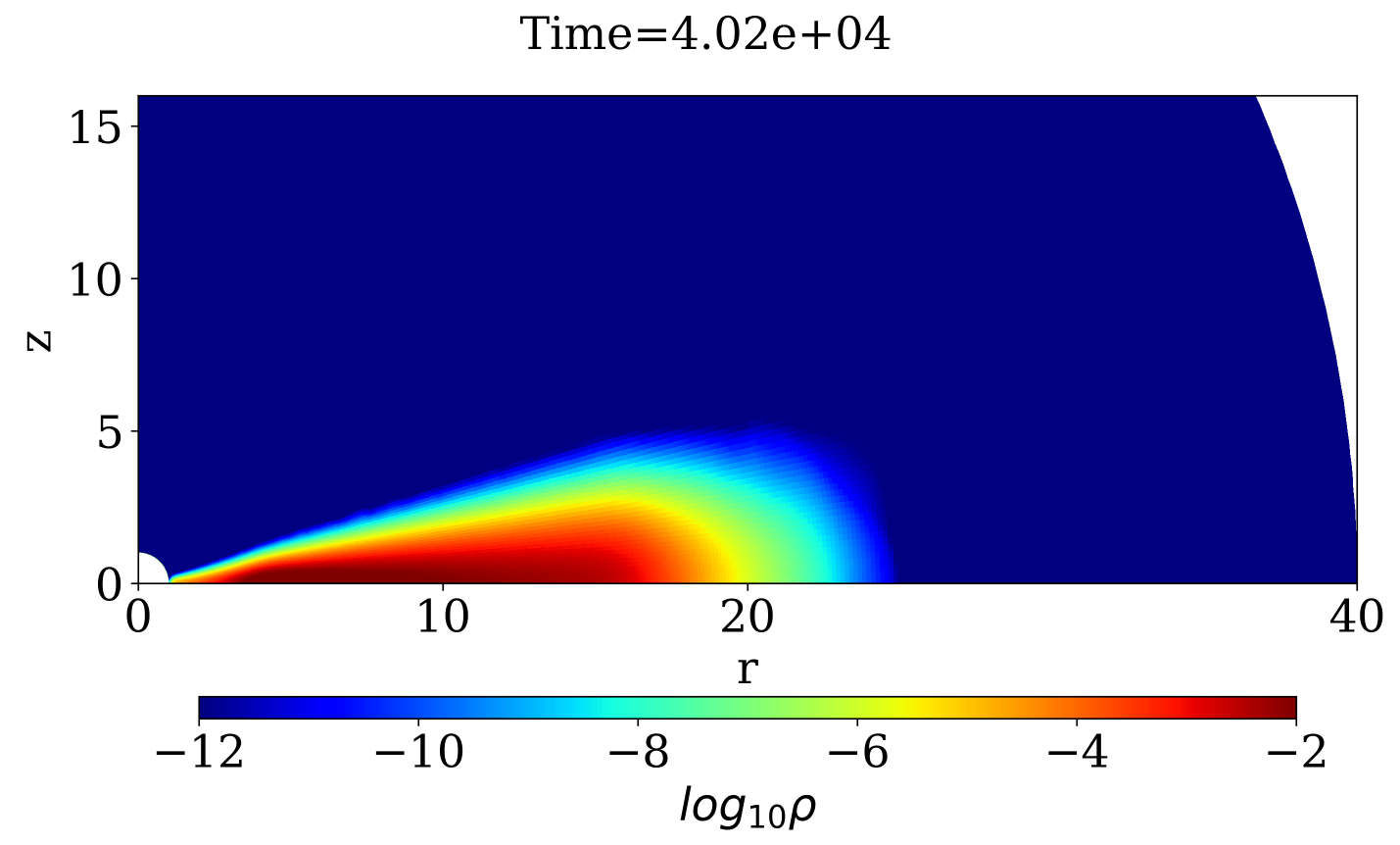}
    \caption{The density of the main body of the disk at different times. Left: the initial density condition of the isolated disk. Right: the density at $t=6400P_{\rm{in}}$.}
    \label{fig:disk2d}
\end{figure*}

\begin{figure}
    \centering
    \includegraphics[width=\columnwidth]{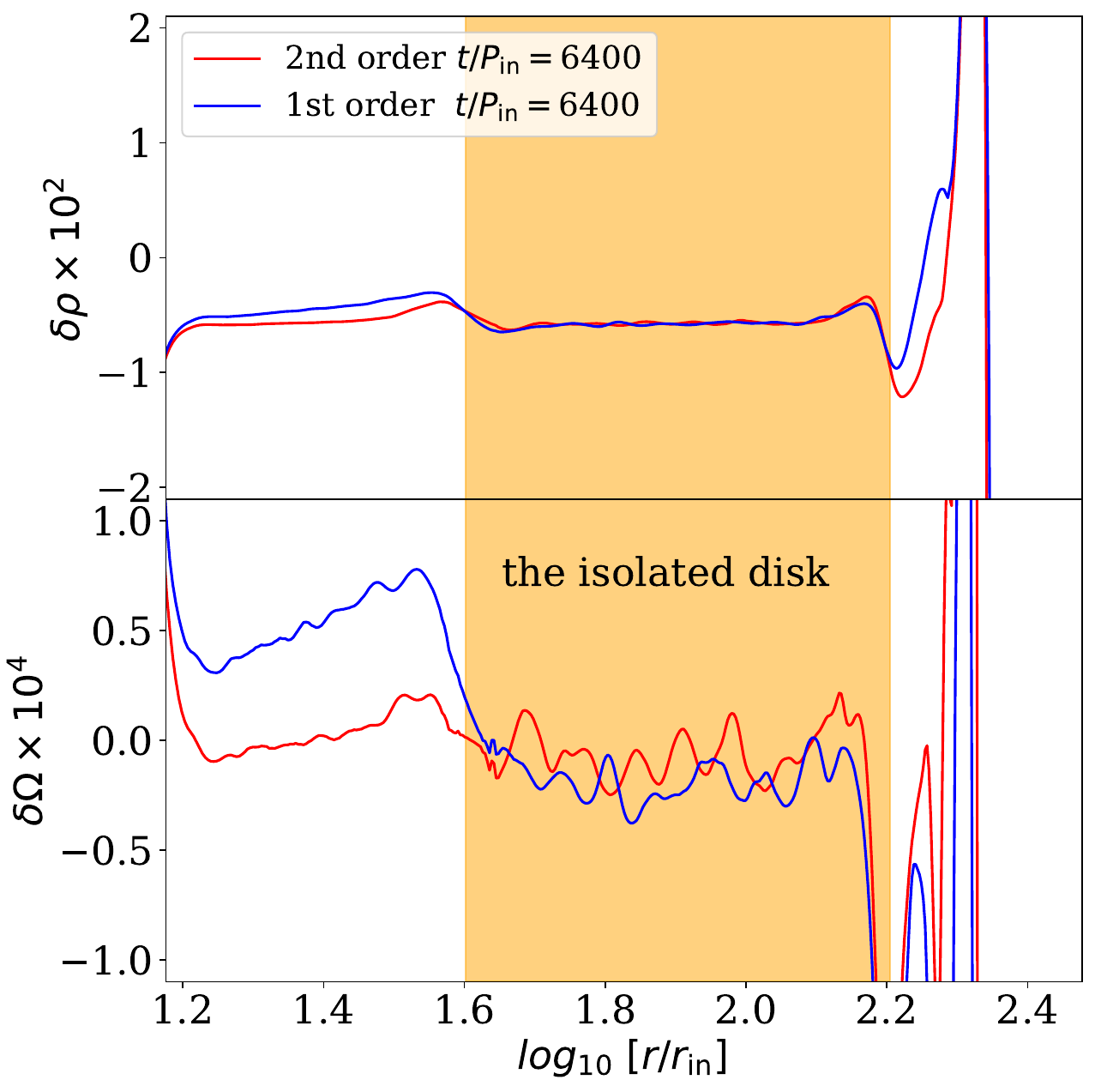}
    \caption{The midplane profiles of the isolated disk, using first order and second order source terms (Appendix \ref{app:sourceterms}) at $t=6400P_{\rm{in}}$, including the steady state solutions in dashed black lines. The orange region is the main body of the isolated disk.}
    \label{fig:diskprofiles}
\end{figure}

\begin{figure}
    \centering
    \includegraphics[width=\columnwidth]{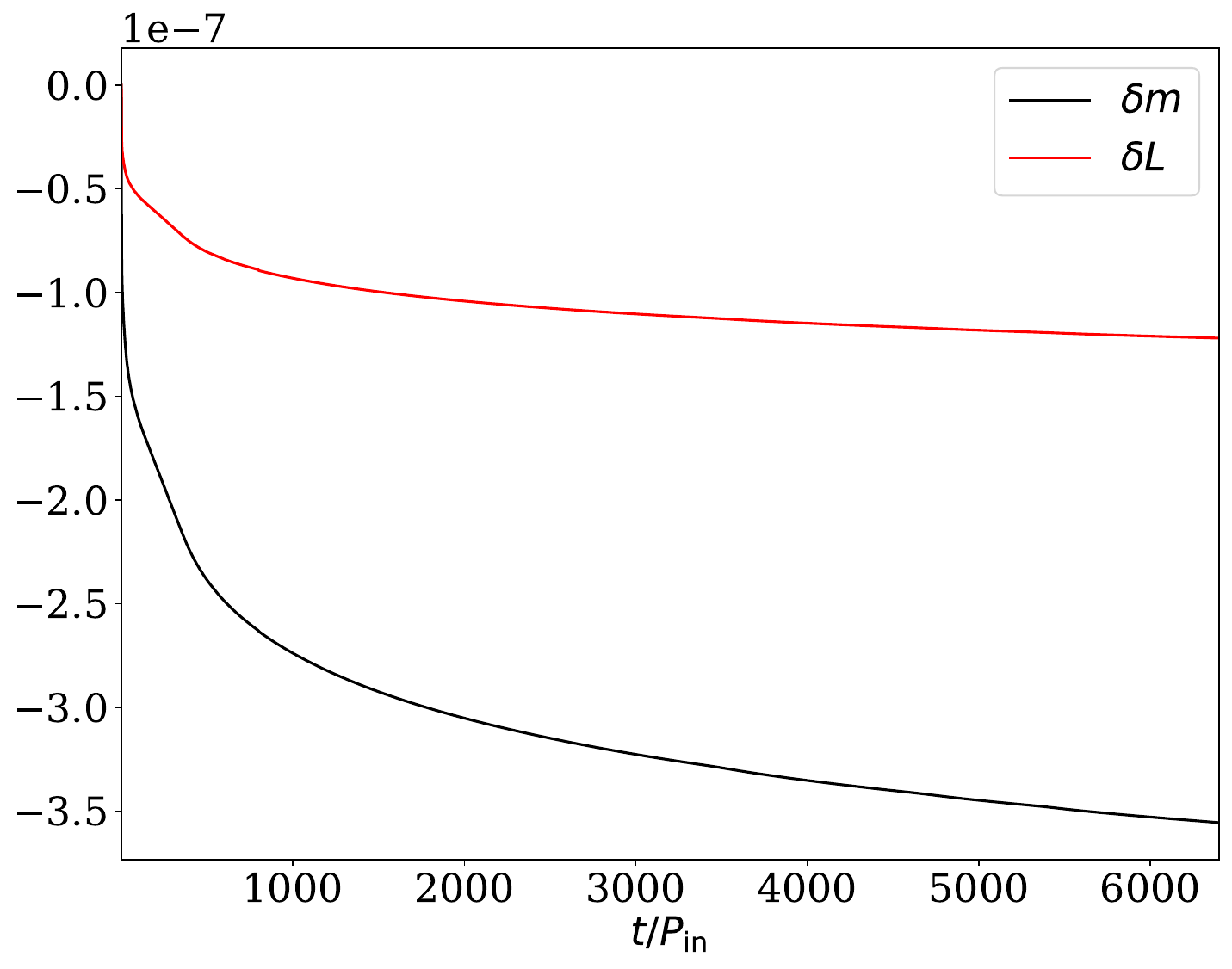}
    \caption{The relative change in mass and angular momentum in the computational domain with the second order scheme (Appendix \ref{app:sourceterms}).}
    \label{fig:diskevolve}
\end{figure}

We evolve the simulation for a total duration of $t = 6,400\,P_{\text{in}}$, where $P_{\text{in}} = 2\pi$ corresponds to the Keplerian orbital period at the reference radius $r_0$. This timespan is equivalent to 800 orbital periods at the inner edge ($r_{\text{in}}$) of the isolated disk, providing a rigorous test of the code's ability to maintain hydrostatic equilibrium over secular timescales. Figure \ref{fig:disk2d} shows the initial density condition and the final density state of the disk. The two profiles remain remarkably similar, with the only discernible differences occurring in the extremely low-density regions. These minor deviations are attributed to the continuous infall of ambient gas from the outer boundary. Furthermore, we evaluate the impact of the source term integration accuracy by comparing simulations using the first- and second-order integrators described in Appendix \ref{app:sourceterms}. The relative differences between these two approaches are illustrated in Figure \ref{fig:diskprofiles}, defined as:
\begin{eqnarray}
    \delta\rho&=&\frac{\rho_{\text{final}}-\rho_{\text{init}}}{\rho_{\text{init}}},    \\
    \delta\Omega&=&\frac{\Omega_{\text{final}}-\Omega_{\text{init}}}{\Omega_{\text{init}}},
\end{eqnarray}
of the midplane profiles at $t=6400P_{\text{in}}$, where ``init" in the subscript means the initial state. We can see that both the first and second order source term integrators can maintain the isolated disk well. The second order source term integrator has smaller variation in $\Omega$.

Figure \ref{fig:diskevolve} illustrates the relative change in the total mass and angular momentum of the disk, defined as:
\begin{eqnarray}
    \delta m&=&\frac{m_{\text{disk}}}{m_{\text{disk,init}}}-1,   \\
    \delta L&=&\frac{L_{\text{disk}}}{L_{\text{disk,init}}}-1,
\end{eqnarray}
where $m_{\text{disk}}$ and $L_{\text{disk}}$ represent the integrated mass and angular momentum within the computational domain. The relative fluctuations in these quantities are remarkably small, on the order of $10^{-7}$. We observe a relatively rapid adjustment during the initial orbits as the analytical steady state relaxes into a numerical equilibrium and a small fraction of the gas exits through the inner boundary. Following this transient phase, the rate of change becomes extremely slow but remains non-zero. This minor evolution can be attributed to two primary factors: (1) the presence of inherent numerical viscosity which facilitates a slow accretion process, and (2) the continuous inflow of zero-angular-momentum ambient gas from the outer boundary, which eventually traverses the inner boundary and carries away a negligible fraction of the disk mass. Given that the relative change is less than $10^{-10}$ per orbit, the precision of the PSAMA scheme is more than sufficient for high-fidelity simulations of long-term disk evolution.

For benchmarking purposes, we performed the same isolated disk test using the {\tt Athena++} framework. We found that maintaining numerical stability in {\tt Athena++} required a relatively high density floor, $\rho_{\rm{floor}}/\rho_{0} \sim 10^{-10}$, to ensure the calculation proceeded without failure\footnote{In contrast, {\tt Guangqi} does not require an enforced $\rho_{\rm{floor}}$ during the hydrodynamic update.}. Reducing the floor density further resulted in either prohibitively small timesteps or code crashes.

Preliminary investigation suggests that the instability is localized in regions of steep density gradients. This behavior likely stems from the fact that while the Riemann fluxes are modified in the $\phi$-momentum (angular momentum) component to ensure conservation, the energy flux is not adjusted commensurately (see Appendix \ref{app:riemannproblems}). This lack of synchronization can lead to unphysical or negative temperatures within a half-timestep. While a higher $\rho_{\rm{floor}}$ can circumvent these numerical artifacts, it introduces an ambient medium with sufficient inertia to perturb the disk, preventing it from maintaining its initial state over many secular timescales.

\subsection{Scalings and performance}

\subsubsection{The weak scaling of pure hydrodynamics}

\begin{figure}
    \centering
    \includegraphics[width=\columnwidth]{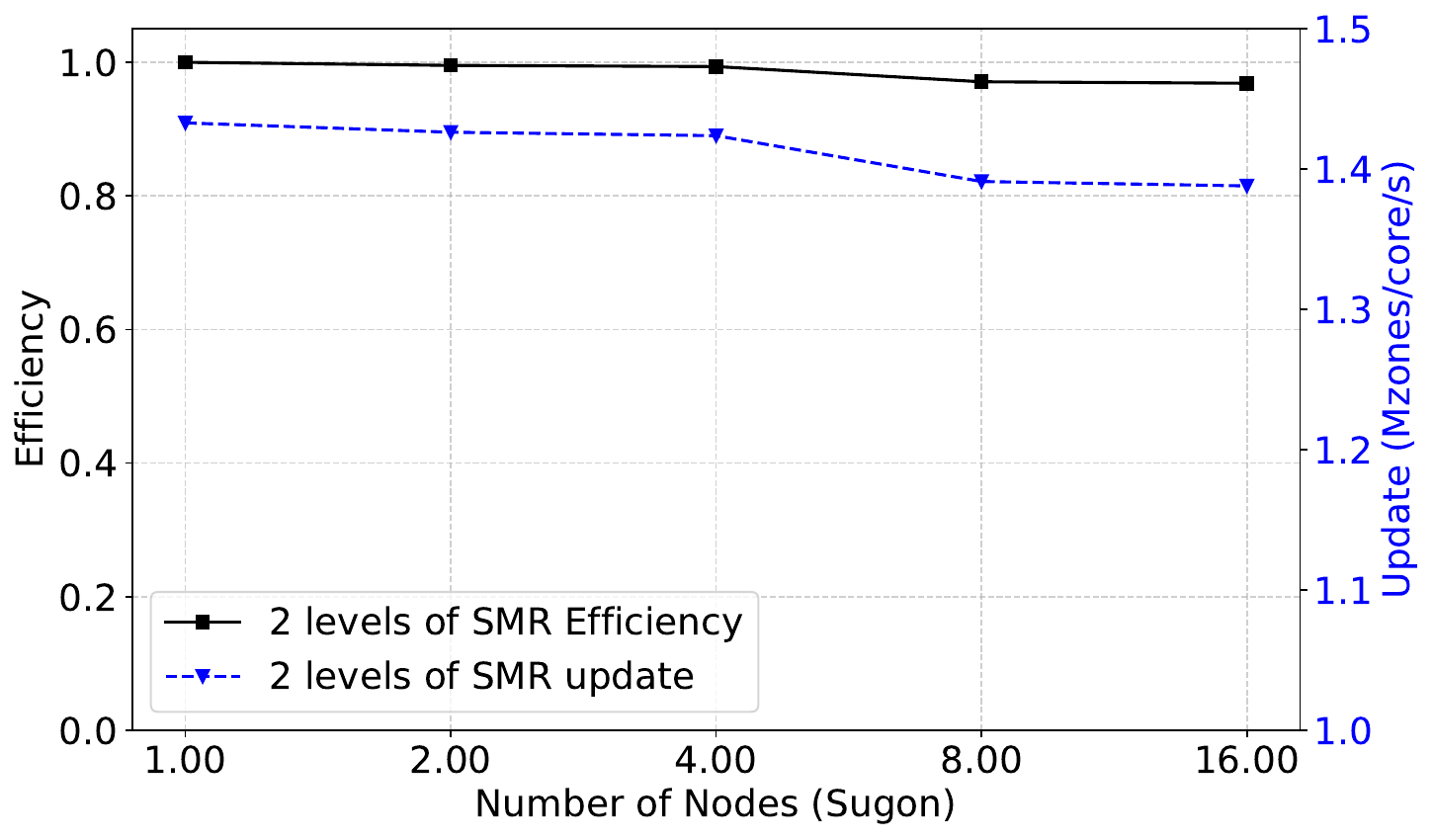}
    \includegraphics[width=\columnwidth]{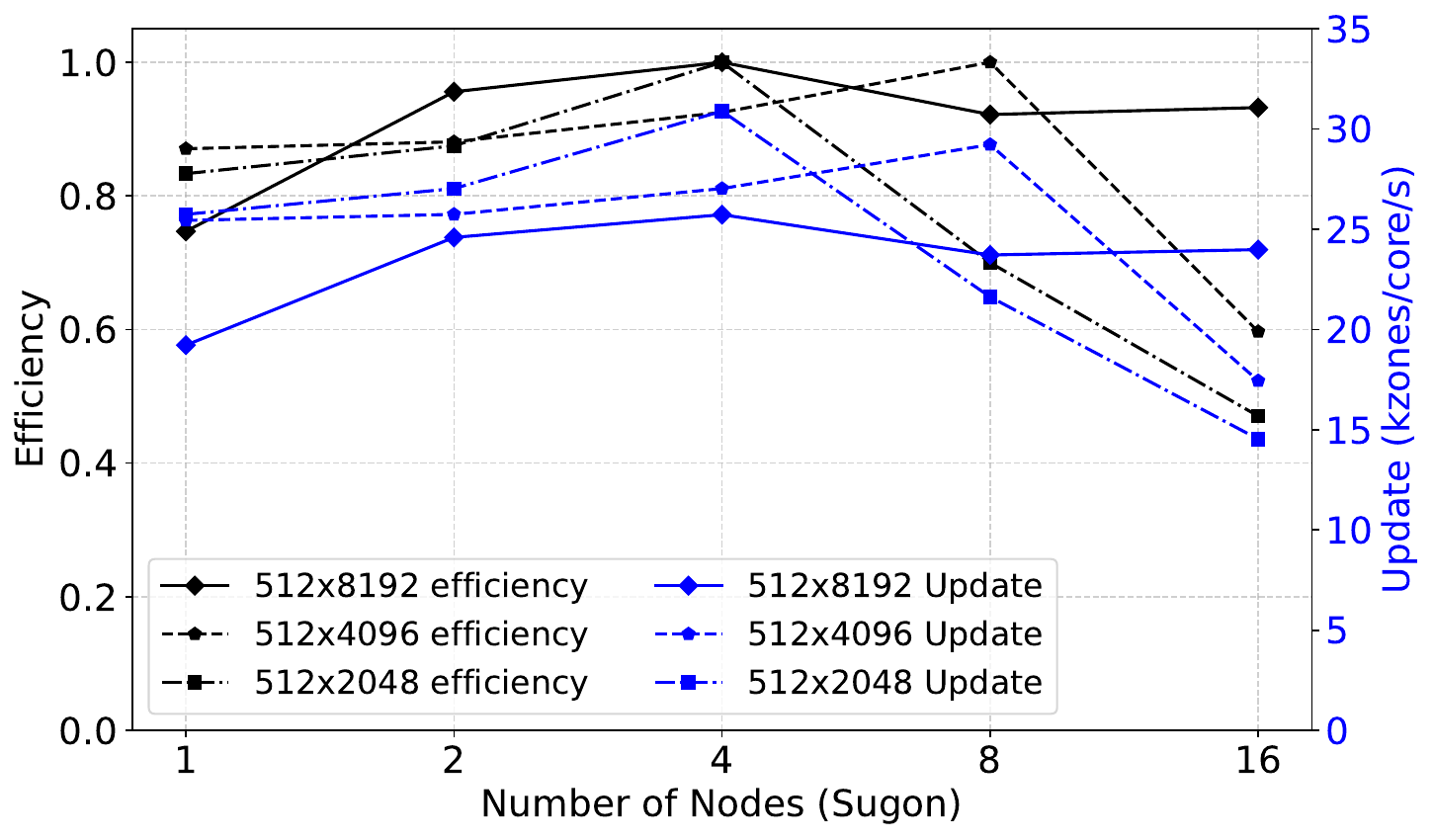}
    \caption{Upper panel: the weak scaling efficiency and performance (Mzones=$10^{6}$ zones) of pure hydrodynamics, maintains above $96\%$ on 16 nodes. Bottom panel: The strong scaling efficiency and performance (kzones=$10^{3}$ zones) of radiation hydrodynamics of {\tt Guangqi}.}
    \label{fig:scaling}
\end{figure}

We evaluate the parallel scaling of {\tt Guangqi} using the KHI simulation described in Section \ref{sec:khi}, employing SMR grid configurations. These scaling tests were performed on a Sugon cluster; each node consists of 2 CPUs with 32 cores and a base clock frequency of 2.5 GHz. Our setup uses a base resolution of $512\times512$ with a block size of $32 \times 32$, which represents a typical configuration for production runs. The two SMR levels are centered on the intervals $y \in [0.245, 0.255]$ and $y \in [-0.255, -0.245]$. To assess weak scaling, we extend the domain in the $x$-direction accordingly. The tests were executed across $[1, 2, 4, 8, 16]$ nodes using 52 cores per node, such that each core processes 16 blocks. The upper panel of Figure \ref{fig:scaling} illustrates the weak scaling efficiency (normalized to a single node) and the corresponding update speed. We find that the weak scaling efficiency remains above $96\%$ up to 16 nodes (832 cores), demonstrating excellent parallel performance in the hydrodynamic part of the code, sufficient for considering more challenging scaling test incorporating radiation.

\subsubsection{The strong scaling of radiation hydrodynamics}

We utilize the supercritical shock setup described in Section \ref{sec:perfectradshock} and extend it to a 2D domain (uniform in the $y$-direction) to evaluate the RHD performance of {\tt Guangqi}. Since iterative solvers are inherently non-linear, we assess the strong scaling of the RHD module across varying total cell counts.

The base resolution in the $x$-coordinate is $n_{x}=512$ over the range $x \in [0, 7 \times 10^{10}]$. In the $y$-coordinate, we test base resolutions of $n_{y} = [2048, 4096, 8192]$ over corresponding ranges of $y_{\text{max}} = [2.8 \times 10^{11}, 5.6 \times 10^{11}, 1.12 \times 10^{12}]$. One level of SMR is applied to the region $x \in [0, 1.75 \times 10^{10}]$. With a fixed block size of $32 \times 32$, the total number of blocks for these cases is $[1792, 3584, 7168]$. Strong scaling tests were conducted on $[1, 2, 4, 8, 16]$ nodes, utilizing 56 cores per node. For these tests, we set $\epsilon_{r}= 10^{-6}$ and perform a single radiation transport calculation per hydrodynamic timestep ($n_{\rm{sub}} = 1$).

The strong scaling efficiencies (normalized to the most efficient node count) and the update speeds are shown in the bottom panel of Figure \ref{fig:scaling}. The efficiency curve exhibits non-linear behavior, which we attribute to the complexity of the iterative solvers. Our results suggest an optimal core count exists for specific problem scales; empirically, we find that a workload of approximately $8 \text{--} 32 \times 10^3$ cells per core is required to maintain high performance. In practice, for instance, our target applications with 2D axisymmetric simulations often require fewer than $10^6$ cells when utilizing SMR and non-uniform grids, making 1--4 nodes sufficient for most production runs. Notably, the RHD calculation is approximately $50 \text{--} 60$ times more computationally expensive than pure hydrodynamics. Given that certain regimes may require subcycling (Section \ref{sec:rhdeos}), optimizing RHD computational efficiency remains a primary objective for future development.

\section{Summary and discussion} \label{sec:disscusions}

In this work, we present {\tt Guangqi}, a new two-dimensional (2D) finite-volume radiation hydrodynamics code. {\tt Guangqi} features a general EoS framework coupled with a fully implicit radiation transport solver under the FLD approximation. It supports block-based SMR/AMR, following the architectural paradigms of codes such as {\tt Athena++} and {\tt Enzo}. The code supports both Cartesian and spherical-polar coordinates with non-uniform grid spacing; notably, our spherical-polar implementation is specifically designed to enhance robustness while strictly conserving angular momentum. Our primary technical highlights include:

\begin{enumerate}
\item General EoS Hydrodynamics: {\tt Guangqi} solves the hydrodynamic equations using the MUSCL scheme integrated with a general EoS HLLC Riemann solver \citep{chen2019} (Section \ref{sec:hydrosolver}). In Section \ref{sec:hydrotest}, we present—for the first time—the combination of AMR and the MUSCL scheme for non-ideal fluids, validating our results against exact general EoS Riemann solutions.
\item Angular Momentum Conservation: In spherical-polar coordinates, the hydrodynamic solver is architected to conserve angular momentum by ensuring compatibility between the Riemann solver and the flux divergence law (Section \ref{sec:angular}). This approach, termed the Passive Scalar Angular Momentum Algorithm (PSAMA), is validated using a specialized isolated disk test in Section \ref{sec:angularmomentumtest}.
\item High-Order Source Terms: To improve accuracy in astrophysical contexts, we implement second-order source terms for geometry and gravity (Appendix \ref{app:sourceterms}). By utilizing the reconstructed slopes from the hydrodynamic solver, these terms are hard-coded into the {\tt Guangqi} pipeline to maximize computational efficiency.
\item Multilevel Implicit Solver: In the presence of SMR/AMR, {\tt Guangqi} solves the multilevel radiation transport and thermodynamic coupling within a single matrix. This unified approach ensures superior energy conservation and consistency across refinement boundaries (Section \ref{sec:implicitamr}).
\end{enumerate}

With its ability to integrate realistic EoS, radiation transport, and the PSAMA formulation, {\tt Guangqi} is well-positioned for a diverse range of applications, particularly involving accretion disks, phase transitions, and optically thick environments. Currently, as a 2D code, {\tt Guangqi} is primarily suited for axisymmetric systems. Future extension to 3D, incorporating the PSAMA framework, will enable more realistic investigations of non-axisymmetric disk dynamics.

Finally, we note that the reliance on the FLD approximation introduces inherent limitations in transition regions between optically thin and thick regimes, as well as in highly anisotropic radiation fields. Furthermore, the computational cost of the implicit solver increases with cell number ($N$) and relative error tolerances ($\epsilon_{r}$). Our experience indicate that the solver runtime scales approximately as $N \log N$, though the dependence on $\epsilon_{r}$ remains complex. A more exhaustive performance analysis and optimization of the implicit framework remain subjects for future research.

\section*{}

The authors would give special thanks to Matthew Knepley and Pierre Jolive for their help with {\tt Petsc} code. We also thank (Alphabetic) Pinghui Huang, Hui Li, Shude Mao, Yuan-sen Ting, Kengo Tomida, Lile Wang, and Weixiao Wang for inspirational discussions. This work is supported by the National Science Foundation of China under grants No. 12103028, 12233004, 12325304, 12473030, 12342501, Tsinghua University Dushi Program and Tsinghua University Initiative Scientific Research Program No. 20233080026. Z.C thank Tsinghua Astrophysics Outstanding Fellowship, and Shuimu Tsinghua Scholar Program for financial support. The Center of High performance computing at Tsinghua University and the National Supercomputer Center in Tianjin provided the computational resources.

\vspace{5mm}
\software{{\tt Matlab} \citep{matlab2018}, {\tt Matplotlib} \citep{hunter2007}, {\tt Petsc} \citep{petsc-efficient,petsc-user-ref}}

\bibliographystyle{aasjournal}
\bibliography{guangqi}

\appendix

\section{Two Riemann problems in a spherical coordinate}\label{app:riemannproblems}

In spherical coordinates, the Euler equations can be formulated using conservative variables based on linear momentum, following conventional numerical approaches. Alternatively, the $\phi$-momentum component can be replaced with angular momentum to improve conservation properties. The compact form of the linear momentum-based hydrodynamic equations (Equations \ref{eqn:spher1}--\ref{eqn:spher5}) at the cell interfaces is expressed as follows:

\begin{equation}\label{eqn:lmbased}
    \pdv{U_\text{lm}}{t}+\pdv{F_\text{lm}}{r}+\pdv{G_\text{lm}}{\theta}=S_\text{lm},
\end{equation}
where, $U_\text{lm},\ F_\text{lm},\ G_\text{lm}$, and $S_\text{lm}$ are the linear momentum based conserved quantities, fluxes, and source terms
\begin{equation}
\begin{split}
U_\text{lm}&=\begin{bmatrix}
\rho	\\
\rho v_r	\\
\rho v_\theta    \\
\rho v_\phi     \\
E
\end{bmatrix},
F_\text{lm}=\begin{bmatrix}
\rho v_r	\\
\rho v_{r}^{2}+p	\\
\rho v_{r}v_{\theta}     \\
\rho v_{r}v_{\phi}      \\
v(E+p)
\end{bmatrix},
G_\text{lm}=\begin{bmatrix}
\rho v_\theta	        \\
\rho v_{r}v_{\theta}	    \\
\rho v_{\theta}^{2}+p    \\
\rho v_{\theta}v_{\phi}     \\
v(E+p)
\end{bmatrix},   \\
S_\text{lm}&=\begin{bmatrix}
0	        \\
\rho g_{r}+2p/r+\rho(v_{\phi}^{2}+v_{\theta}^{2})/r	    \\
[\cot\theta(\rho v_{\phi}^{2}+p)-\rho v_{r}v_{\theta}]/r    \\
-(\rho v_{r}v_{\theta}-\cot\theta\rho v_{\theta}v_{\phi})/r   \\
\rho g_{r} v_{r}
\end{bmatrix}.
\end{split}
\end{equation}

Alternatively, the hydrodynamic equations written based on angular momentum read
\begin{equation}\label{eqn:ambased}
    \pdv{U_\text{am}}{t}+\pdv{F_\text{am}}{r}+\pdv{G_\text{am}}{\theta}=S_\text{am},
\end{equation}
where
\begin{equation}
\begin{split}
U_\text{am}&=\begin{bmatrix}
\rho	\\
\rho v_r	\\
\rho v_\theta    \\
\rho l     \\
E
\end{bmatrix},
F_\text{am}=\begin{bmatrix}
\rho v_r	\\
\rho v_{r}^{2}+p	\\
\rho v_{r}v_{\theta}     \\
\rho v_{r}l      \\
v(E+p)
\end{bmatrix},
G_\text{am}=\begin{bmatrix}
\rho v_\theta	        \\
\rho v_{r}v_{\theta}	    \\
\rho v_{\theta}^{2}+p    \\
\rho v_{\theta}l     \\
v(E+p)
\end{bmatrix},   \\
S_\text{am}&=\begin{bmatrix}
0	        \\
\rho g_{r}+2p/r+\rho[(l/r\sin\theta)^{2}+v_{\theta}^{2}]/r	    \\
\{\cot\theta[\rho(l/r\sin\theta)^{2}+p]-\rho v_{r}v_{\theta}\}/r    \\
0   \\
\rho g_{r} v_{r}
\end{bmatrix}.
\end{split}
\end{equation}

While the left-hand side (LHS) of Equation \ref{eqn:lmbased} can be resolved using a standard Riemann solver \citep{toro2013}, Equation \ref{eqn:ambased} presents a more complex solution structure. This is because the total energy, $E$, is coupled to the angular momentum in a non-trivial manner, as expressed by Equation \ref{eqn:energysplit}. Numerical inconsistency arises when the Riemann solution from the linear-momentum-based Equation \ref{eqn:lmbased} is used to apply the flux divergence law to the angular-momentum-based state vector, $U_\text{am}$.

Ideally, one would solve the Riemann problem directly for Equation \ref{eqn:ambased} and apply the flux divergence law to $U_\text{am}$ to ensure the simultaneous and consistent conservation of both angular momentum and energy. However, given the significant analytical complexity of this approach, we instead propose the passive scalar angular momentum algorithm (PSAMA) in Section \ref{sec:angular}. This method prioritizes a more consistent angular momentum conservation formula to provide enhanced numerical robustness and stability, albeit at the cost of precise energy conservation---which is already subject to small deviations due to the iterative solvers used for radiation transport and gravitational source terms.

\section{Derivation of PSAMA}\label{app:derivation}

After splitting the energy into two parts in Equation \ref{eqn:energysplit}, we adapt Equation \ref{eqn:spher5} correspondingly,
\begin{eqnarray}
    \pdv{E'}{t}+\frac{1}{r^{2}}\pdv{[r^{2}(E'+p) v_{r}]}{r}+\frac{1}{r\sin\theta}\pdv{[\sin\theta(E'+p)v_{\theta}]}{\theta}\nonumber&\\
    +\pdv{e_{k\phi}}{t}+\nabla\cdot({e_{k\phi}\vec{v}})=\rho g_{r}v_{r}.\quad\quad&
\end{eqnarray}
The key is to express ${\partial e_{k\phi}}/{\partial t}+\nabla\cdot({e_{k\phi}\vec{v}})$ in terms of the known variables, the derivation is the following,
\begin{equation}
    2\pdv{e_{k\phi}}{t}=\pdv{(\rho\Omega l)}{t}=\Omega\pdv{\rho l}{t}+\rho l\pdv{\Omega}{t},  \label{eqn:ekphi1}
\end{equation}
where $\Omega$ is the angular frequency around the $z$ axis. Apparently, from Equation \ref{eqn:spher4am},
\begin{equation}
    \Omega\pdv{\rho l}{t}+\Omega\nabla\cdot{(\rho\vec{v}l)}=0.\label{eqn:omega1}
\end{equation}
On the other hand, using continuity equation and Equation \ref{eqn:spher4am}, we can obtain,
\begin{equation}
    \rho\pdv{l}{t}+\rho\vec{v}\cdot\nabla l=0.
\end{equation}
Substitute $l=\Omega(r\sin\theta)^2$ and multiply by $\Omega$, the above equation becomes,
\begin{equation}
    \rho l\pdv{\Omega}{t}+\rho\Omega\vec{v}\cdot\nabla l=0.\label{eqn:omega2}
\end{equation}
Adding Equation \ref{eqn:omega1} and \ref{eqn:omega2}, we find,
\begin{equation}
    2\pdv{e_{k\phi}}{t}=-\Omega[\rho\vec{v}\cdot\nabla l+\nabla\cdot{(\rho\vec{v}l)}]. \label{eqn:ekphi2}
\end{equation}
Meanwhile,
\begin{eqnarray}
    \nabla l&=&[(r\sin\theta)^{2}\pdv{\Omega}{r}+2r(\sin\theta)^{2}\Omega,\nonumber\\
    &&r(\sin\theta)^{2}\pdv{\Omega}{\theta}+2r\sin\theta\cos\theta\Omega,0]^{T} \nonumber\\
    &=&(r\sin\theta)^{2}\nabla\Omega+2r\sin\theta\Omega[\sin\theta,\cos\theta,0]^{T}    \nonumber\\
    &=&\frac{l}{\Omega}\nabla\Omega+2\vec{v}_{\Omega}, \label{eqn:nablal}
\end{eqnarray}
where $\vec{v}_{\Omega}=r\sin\theta\Omega[\sin\theta,\cos\theta,0]^{T}$ is in principle similar to the origin of centrifugal force. Next, substitute Equation \ref{eqn:nablal} into \ref{eqn:ekphi2},
\begin{eqnarray}
    2\pdv{e_{k\phi}}{t}&=&-\Omega\rho\vec{v}\cdot\bigg[\frac{l}{\Omega}\nabla\Omega+2\vec{v}_{\Omega}\bigg]-\Omega\nabla\cdot{(\rho\vec{v}l)}   \nonumber\\
    &=&-\nabla\cdot{(\rho\vec{v}\Omega l)}-2\rho\Omega\vec{v}_{\Omega}\cdot\vec{v}\nonumber\\
    &=&-\nabla\cdot{(\rho\vec{v}\Omega l)}-2\rho\vec{f}_{\Omega}\cdot\vec{v},\label{eqn:ekphi3}
\end{eqnarray}
where $\vec{f}_{\Omega}=\Omega\vec{v}_{\Omega}=[v_{\phi}^2/r,v_{\phi}^2\cot\theta/r,0]$ is the centrifugal force. Therefore, reorganizing Equation \ref{eqn:ekphi3}, we obtain
\begin{equation}
    \pdv{e_{k\phi}}{t}+\nabla\cdot{(e_{k\phi}\vec{v})}=-\rho\vec{f}_{\Omega}\cdot\vec{v}.
\end{equation}
Not surprisingly,
\begin{equation}
    \rho\vec{v}\cdot\vec{f}_{\Omega}=\rho\vec{v}\cdot\bigg[\frac{v_{\phi}^2}{r},\frac{v_{\phi}^2\cot\theta}{r},0\bigg]^{T},
\end{equation}
is the work done by the centrifugal force.

\section{Second order integration of the source terms}\label{app:sourceterms}

The Euler equations in a spherical coordinate will give rise to geometric source terms (see Equation \ref{eqn:spher1}-\ref{eqn:spher5}), including $2p/r$, $\rho v_{\phi}^2/r$, $p\cot\theta/r$, etc. Each of them may become important in different regions, for example, $2p/r$ or $\rho v_{\phi}^2/r$ may balance the gravity in the radial direction in an atmosphere or an accretion disk, respectively. \citet{mignone2014} took a general approach to the source terms which may increase the complexity of the code. On the other hand, well-balanced scheme takes an assumed steady state solution into account \citep{kappeli2022} that can cancel out most of the truncation error if the fluid is indeed close to the assumed steady state. In {\tt Guangqi}, we hard code the source terms that are most relevant in astrophysics (geometric source terms and central gravity) and do not presume any steady state solutions. Because in many cases, we won't know which region is in which steady state.

The second order integration of the source term is done by making use of the slope limited hydrodynamic variables in a cell's volume integration. Specifically, the pressure source term in the spherical coordinate in $\{i,j\}$ cell is
\begin{equation}\label{eqn:pressource}
    \int_{V}\frac{2p}{r}r^{2}\sin\theta drd\theta=\int_{V}2pr\sin\theta drd\theta,
\end{equation}
where,
\begin{eqnarray}
    p=p_{i,j}+\frac{\partial p}{\partial r}(r-r_{i,j})+\frac{\partial p}{\partial\theta}(\theta-\theta_{i,j})
\end{eqnarray}
where, $p_{i,j}$ is the average pressure of the cell, $\pdv{p}{r}$ and $\pdv{p}{\theta}$ are the slopes that are also used in the reconstruction step in Section \ref{sec:hydrosolver}, and $r_{i,j}$ and $\theta_{i,j}$ are the volume centric coordinate of the cell. When taking $\partial p/\partial r=\partial p/\partial\theta=0$, the source term reduces to the first order case. The final expression of \ref{eqn:pressource} is hard coded in {\tt Guangqi}. In the same manner, we show how we calculate the gravitational work done to the gas in a spherical coordinate
\begin{equation}
    \int_{V}\frac{\rho v_{r}}{r^{2}}r^{2}\sin\theta drd\theta=\int_{V}\rho v_{r}r^{2}\sin\theta drd\theta,
\end{equation}
where,
\begin{eqnarray}
    \rho&=&\rho_{i,j}+\frac{\partial\rho}{\partial r}(r-r_{i,j})+\frac{\partial\rho}{\partial\theta}(\theta-\theta_{i,j}),   \\
    v_{r}&=&v_{r,i,j}+\frac{\partial v_{r}}{\partial r}(r-r_{i,j}),
\end{eqnarray}
where we omit the $\theta$ dependence in $v_{r}$ for simplicity. The final expression also omit the product of the terms such as $\partial\rho/\partial r\cdot \partial v_{r}/\partial r$ as they are high order terms.

By hard coding these source terms in {\tt Guangqi}, we can maintain the hydrodynamic solver second order in space and time in the spherical coordinate without significantly increasing the computational cost. Section \ref{sec:angularmomentumtest} has a steady state disk test that includes all the source terms and we show that the disk is maintained steady in {\tt Guangqi} for thousands of orbits.

\end{CJK*}
\end{document}